\documentclass[aps,prd,preprint,a4paper,showpacs]{revtex4}
\usepackage{graphicx,natbib}
\usepackage{amsmath}
\newcommand{\bi}[1]{\mbox{\boldmath ${#1}$}}
\newcommand{\dis}{\displaystyle}
\begin{document}
\title{Radiative Decays Involving $f_0(980)$ and $a_0(980)$ \\
and \\
Mixing Between Low and High Mass Scalar Mesons }
\date{\today}
\vspace*{1cm}
\author{T. Teshima}
\email{teshima@isc.chubu.ac.jp}
\author{I. Kitamura}
\author{N. Morisita}
\affiliation{Department of Applied Physics,  Chubu University, Kasugai 
487-8501, Japan}
\begin{abstract}
We analyze the experimental data for $\phi\to f_0(980)\gamma$, $\phi\to a_0(980)
\gamma$, $f_0(980)\to\gamma\gamma$ and $a_0(980)\to\gamma\gamma$ decay widths in 
a framework where $f_0(980)$ and $a_0(980)$ are assumed to be mainly 
$qq\bar{q}\bar{q}$ low mass scalar mesons and mixed with $q\bar{q}$ high mass 
scalar mesons. 
Applied the vector meson dominance model (VDM), these decay amplitudes are 
expressed by coupling parameters $B$ describing the $S$($qq\bar{q}\bar{q}$ scalar 
meson)-V(vector meson)-V(vector meson) coupling and $B'$ describing the 
$S'$($q\bar{q}$ scalar meson)-V-V coupling. 
Adopting the magnitudes for $B$ and $B'$ as $\sim2.8{\rm GeV^{-1}}$ and 
$\sim12{\rm GeV^{-1}}$, respectively, the mixing angle between $a_0(980)$ and 
$a_0(1450)$ as $\sim9^\circ$, and the mixing parameter $\lambda_{01}$ causing the 
mixing between $I=0$ $qq\bar{q}\bar{q}$ state and $q\bar{q}$ state as $\sim 
0.24{\rm GeV}^2$ , we can interpret these experimental data, consistently.
\end{abstract}
\pacs{12.15.Ff, 13.15.+g, 14.60.Pq}
\preprint{CU-TP/05-01}
\maketitle
%\setlength{\baselineskip}{0.33in}
%%%%%%%%%%%%% section 1%%%%%%%%%%%%%%%
\section{Introduction}
From the recent experimental and theoretical analyses for the $f_0(600)$ and 
$\kappa(900)$, these scalar mesons are considered as the light scalar nonet 
together with the $f_0(980)$ and $a_0(980)$ \cite{sigma-kappa}. 
From the status that though the masses of $f_0(980)$ and $a_0(980)$ are degenerate, 
the $f_0(980)$ state has the strangeness flavor rich character,  many authors 
suggest this nonet as non $q\bar{q}$ structure, e. g. $K\bar{K}$ molecule 
\cite{molecule}, $qq\bar{q}\bar{q}$ state \cite{jaffe}. 
On this standpoint, many literatures assume that the higher mass scalar mesons, 
$f_0(1370)$, $a_0(1450)$, $K_0^*(1430)$, $f_0(1500)$ and $f_0(1710)$ would be 
traditional $q\bar{q}$ nonet and glueball state. 
\par
In order to confirm the structure of the light scalar mesons, that is whether the 
light scalar mesons are constituted of $q\bar{q}$ or $qq\bar{q}\bar{q}$, 
the radiative decays of the $\phi$ meson to the scalar mesons $f_0(980)$ and 
$a_0(980)$, $\phi\to f_0\gamma$ and $\phi\to a_0\gamma$, have long been analyzed by 
assuming the model, in which the decay $\phi\to f_0(a_0)\gamma$ proceeds 
through the charged $K$ loop, $\phi\to K^+K^-\to f_0(a_0)\gamma$ \cite{phi decay I}. Almost literature suggests that $f_0(980)$ and $a_0(980)$ mesons contain 
significant $qq\bar{q}\bar{q}$ content, specifically being $(u\bar{u}\pm b\bar{b})
s\bar{s}$. 
\par
The two $\gamma$ decays of the $f_0(980)$ and $a_0(980)$ mesons are also analyzed 
by using the various models, the linear sigma model \cite{2gamma.1}, 
vector meson dominance model (VDM) \cite{phi decay II} and the quark-hadron 
duality idea \cite{2gamma.2}.
Literature \cite{2gamma.1} suggested that the $f_0(980)$ meson 
is mostly composed of $s\bar{s}$ component under the picture of the nonet 
scalar for light scalar mesons $f_0(600), \kappa(900), a_0(980)$ and 
$f_0(980)$.  
Literature \cite{phi decay II} studied the $a_0(980)\to\gamma\gamma$ and $f_0(980)
\to\gamma\gamma$ decays and further $\phi\to f_0(980)\gamma$ and $\phi\to 
a_0(980)\gamma$ decays comprehensively assuming the VDM and the $qq\bar{q}\bar{q}$ 
structure for light scalar mesons. 
It could explain the experimental results for radiative decays except for the 
$\phi\to f_0(980)\gamma$ decay.  
Literature \cite{2gamma.2} considered the nonet scalar for light scalar mesons 
and assumed that these were composed of $q\bar{q}$ or $qq\bar{q}\bar{q}$ states. 
The author analyzed the two $\gamma$ decays considering the mixing between 
$qq\bar{q}\bar{q}$ and $q\bar{q}$, latter of which is the dominant structure 
of the higher mass scalar mesons . 
\par
We analyzed the mixing between the light scalar nonet, $f_0(600)$, 
$\kappa(900)$, $a_0(980)$, $f_0(980)$ assumed as $qq\bar{q}\bar{q}$ states 
dominantly and high mass scalar nonet + glueball, $f_0(1370)$, $a_0(1450)$, 
$K_0^*(1430)$, $f_0(1500)$, $f_0(1710)$ assumed as $L=1$ $q\bar{q}$ states 
dominantly + glueball state, in our previous work \cite{teshima1}. 
The estimated mixing is very strong because of the fact that the high mass scalar 
meson masses are very high compared with the masses supposed from the $L=1$ 
$q\bar{q}$ $1^{++}$ and $2^{++}$ meson masses and relation $m^2(2^{++})-m^2(1^{++})
=2(m^2(1^{++})-m^2(0^{++}))$ resulting from the $L\cdot S$ force.    
Literature \cite{schechter} took the similar conclusion to ours in the mixing 
for $I=1$ mesons and $I=1/2$ mesons. 
That the mixing between $qq\bar{q}\bar{q}$ state and $q\bar{q}$ state is strong is 
recognized from the fact that the transition between $qq\bar{q}\bar{q}$ and 
$q\bar{q}$ states is caused by the OZI rule allowed diagram. 
In next work \cite{teshima2}, we pursued this problem analyzing the decay 
processes in which light scalar mesons and high mass scalar meson decays to two 
pseudoscalar mesons, and get the result that the mixing angle between $I=1$ 
$a_0(980)$ and $a_0(1450)$ is $\sim 10^\circ$. 
\par
In the present work, we wil analyze the $\phi(1020)\to a_0(980)\gamma$, $\phi(1020)
\to f_0(980)\gamma$, $a_0(980)\to\gamma\gamma$ and $f_0(980)\to\gamma\gamma$ decays 
assuming that the light scalar mesons have the $qq\bar{q}\bar{q}$ component and 
$q\bar{q}$ component, and we will reveal the mixing ratio of these components. 
We analyze this problem using the vector dominance model, wherein we can treat 
these radiative decay processes comprehensively.  
%%%%%%%%%%% section 2 %%%%%%%%%%%%%%
\section{Mixing between Low and High Mass Scalar Mesons}
In this section, we briefly review the mixing among the low mass scalar, high 
mass scalar and glueball discussed in our previous work \cite{teshima1, teshima2}.
The $qq\bar{q}\bar{q}$ scalar $SU(3)$ nonet $S^b_a$ are represented by the quark 
triplet $q_a$ and anti-quark triplet $\bar{q^a}$ as \cite{jaffe}, \cite{schechter}
\begin{equation}
S^a_b\sim\epsilon^{acd}q_cq_d\epsilon_{bef}\bar{q}^e\bar{q}^f
\end{equation}
and have the following flavor configuration:
\begin{equation}
    \begin{array}{ccc}
    \bar{s}\bar{d}us,\ \frac{1}{\sqrt{2}}(\bar{s}\bar{d}ds-\bar{s}\bar{u}us),\ 
    \bar{s}\bar{u}ds&\Longleftrightarrow&a_0^+,\ a_0^0,\ a_0^-\\
    \bar{s}\bar{d}ud,\ \bar{s}\bar{u}ud,\ \bar{u}\bar{d}us,\ \bar{u}\bar{d}ds
    &\Longleftrightarrow&\kappa^+,\ \kappa^0,\ \overline{\kappa^0},\ \kappa^-\\
    \frac{1}{\sqrt{2}}(\bar{s}\bar{d}ds+\bar{s}\bar{u}us)&\Longleftrightarrow&f_{NS}
    \sim f_0(980)\\
    \bar{u}\bar{d}ud&\Longleftrightarrow&f_{NN}\sim f_0(600) 
    \end{array} \nonumber
\end{equation}
We use the notation $f_{NS}$ for $\frac{1}{\sqrt{2}}(\bar{s}\bar{d}ds+\bar{s}
\bar{u}us)$ and $f_{NN}$ for $\bar{u}\bar{d}ud$ in this paper, but we used $f_{N}$ 
and $f_{S}$ for $\frac{1}{\sqrt{2}}(\bar{s}\bar{d}ds+\bar{s}\bar{u}us)$ and 
$\bar{u}\bar{d}ud$, respectively in our previous literature \cite{teshima1, 
teshima2}.
The high mass scalar mesons $S'^a_b$ are the ordinary $SU(3)$ nonet 
$$ S'^a_b\sim \bar{q}^aq_b. $$
%%%%%%%%%%%%%%%%%%%%%(2-1)%%%%%%%%%%%%%%%%%%%%%%%%
\subsection{Inter-mixing between $I=1, 1/2$ low and high mass scalar mesons} 
The mixing between $qq\bar{q}\bar{q}$ and $q\bar{q}$ states, for which we 
call "inter-mixing", may be large, because the transition between $qq\bar{q}
\bar{q}$ and $q\bar{q}$ states is caused by the OZI rule allowed diagram shown 
in fig. 1.
%%%%%%%%%%%%%%%%%%%%%%%%%%%%%%%%%%%
\begin{figure}
\vspace{0.5cm}
\begin{center}
%WinTpicVersion3.08
\unitlength 0.1in
\begin{picture}( 24.5000,  6.1000)(  2.4000, -8.0000)
% LINE 1 0 3 0
% 8 600 200 2600 200 2600 800 600 800 600 600 1500 600 1500 400 600 400
% 
\special{pn 13}%
\special{pa 600 200}%
\special{pa 2600 200}%
\special{fp}%
\special{pa 2600 800}%
\special{pa 600 800}%
\special{fp}%
\special{pa 600 600}%
\special{pa 1500 600}%
\special{fp}%
\special{pa 1500 400}%
\special{pa 600 400}%
\special{fp}%
% CIRCLE 1 0 3 0
% 4 1500 500 1500 600 1500 600 1500 400
% 
\special{pn 13}%
\special{ar 1500 500 100 100  4.7123890 6.2831853}%
\special{ar 1500 500 100 100  0.0000000 1.5707963}%
% LINE 2 0 3 0
% 4 1300 200 1300 800 1900 800 1900 200
% 
\special{pn 8}%
\special{pa 1300 200}%
\special{pa 1300 800}%
\special{fp}%
\special{pa 1900 800}%
\special{pa 1900 200}%
\special{fp}%
% BOX 2 5 3 0
% 2 1900 200 2610 800
% 
\special{pn 8}%
\special{pa 1900 200}%
\special{pa 2610 200}%
\special{pa 2610 800}%
\special{pa 1900 800}%
\special{pa 1900 200}%
\special{ip}%
% LINE 3 0 3 0
% 20 2600 200 2010 790 2480 200 1900 780 2360 200 1900 660 2240 200 1900 540 2120 200 1900 420 2000 200 1900 300 2610 310 2130 790 2610 430 2250 790 2610 550 2370 790 2610 670 2490 790
% 
\special{pn 4}%
\special{pa 2600 200}%
\special{pa 2010 790}%
\special{fp}%
\special{pa 2480 200}%
\special{pa 1900 780}%
\special{fp}%
\special{pa 2360 200}%
\special{pa 1900 660}%
\special{fp}%
\special{pa 2240 200}%
\special{pa 1900 540}%
\special{fp}%
\special{pa 2120 200}%
\special{pa 1900 420}%
\special{fp}%
\special{pa 2000 200}%
\special{pa 1900 300}%
\special{fp}%
\special{pa 2610 310}%
\special{pa 2130 790}%
\special{fp}%
\special{pa 2610 430}%
\special{pa 2250 790}%
\special{fp}%
\special{pa 2610 550}%
\special{pa 2370 790}%
\special{fp}%
\special{pa 2610 670}%
\special{pa 2490 790}%
\special{fp}%
% BOX 2 5 3 0
% 2 590 190 1300 790
% 
\special{pn 8}%
\special{pa 590 190}%
\special{pa 1300 190}%
\special{pa 1300 790}%
\special{pa 590 790}%
\special{pa 590 190}%
\special{ip}%
% LINE 3 0 3 0
% 14 1120 600 930 790 1000 600 810 790 880 600 690 790 760 600 590 770 640 600 590 650 1240 600 1050 790 1300 660 1170 790
% 
\special{pn 4}%
\special{pa 1120 600}%
\special{pa 930 790}%
\special{fp}%
\special{pa 1000 600}%
\special{pa 810 790}%
\special{fp}%
\special{pa 880 600}%
\special{pa 690 790}%
\special{fp}%
\special{pa 760 600}%
\special{pa 590 770}%
\special{fp}%
\special{pa 640 600}%
\special{pa 590 650}%
\special{fp}%
\special{pa 1240 600}%
\special{pa 1050 790}%
\special{fp}%
\special{pa 1300 660}%
\special{pa 1170 790}%
\special{fp}%
% LINE 3 0 3 0
% 14 1080 400 890 590 960 400 770 590 840 400 650 590 720 400 590 530 1200 400 1010 590 1300 420 1130 590 1300 540 1250 590
% 
\special{pn 4}%
\special{pa 1080 400}%
\special{pa 890 590}%
\special{fp}%
\special{pa 960 400}%
\special{pa 770 590}%
\special{fp}%
\special{pa 840 400}%
\special{pa 650 590}%
\special{fp}%
\special{pa 720 400}%
\special{pa 590 530}%
\special{fp}%
\special{pa 1200 400}%
\special{pa 1010 590}%
\special{fp}%
\special{pa 1300 420}%
\special{pa 1130 590}%
\special{fp}%
\special{pa 1300 540}%
\special{pa 1250 590}%
\special{fp}%
% LINE 3 0 3 0
% 14 1040 200 850 390 920 200 730 390 800 200 610 390 680 200 590 290 1160 200 970 390 1280 200 1090 390 1300 300 1210 390
% 
\special{pn 4}%
\special{pa 1040 200}%
\special{pa 850 390}%
\special{fp}%
\special{pa 920 200}%
\special{pa 730 390}%
\special{fp}%
\special{pa 800 200}%
\special{pa 610 390}%
\special{fp}%
\special{pa 680 200}%
\special{pa 590 290}%
\special{fp}%
\special{pa 1160 200}%
\special{pa 970 390}%
\special{fp}%
\special{pa 1280 200}%
\special{pa 1090 390}%
\special{fp}%
\special{pa 1300 300}%
\special{pa 1210 390}%
\special{fp}%
% FUNC 2 0 3 0
% 9 1580 500 1900 500 1750 490 1800 500 1750 450 1580 500 1900 500 0 2 0 0
% sint
\special{pn 8}%
% FUNC 1 0 3 0
% 9 1300 200 1900 800 1600 500 1800 500 1600 300 1600 200 1900 800 0 5 0 0
% 0.1sin(10x)
\special{pn 13}%
\special{pa 1300 514}%
\special{pa 1306 516}%
\special{pa 1310 520}%
\special{pa 1316 520}%
\special{pa 1320 520}%
\special{pa 1326 520}%
\special{pa 1330 516}%
\special{pa 1336 514}%
\special{pa 1340 508}%
\special{pa 1346 504}%
\special{pa 1350 500}%
\special{pa 1356 494}%
\special{pa 1360 490}%
\special{pa 1366 486}%
\special{pa 1370 482}%
\special{pa 1376 482}%
\special{pa 1380 480}%
\special{pa 1386 482}%
\special{pa 1390 482}%
\special{pa 1396 486}%
\special{pa 1400 490}%
\special{pa 1406 494}%
\special{pa 1410 498}%
\special{pa 1416 504}%
\special{pa 1420 508}%
\special{pa 1426 512}%
\special{pa 1430 516}%
\special{pa 1436 518}%
\special{pa 1440 520}%
\special{pa 1446 520}%
\special{pa 1450 520}%
\special{pa 1456 516}%
\special{pa 1460 514}%
\special{pa 1466 510}%
\special{pa 1470 504}%
\special{pa 1476 500}%
\special{pa 1480 494}%
\special{pa 1486 490}%
\special{pa 1490 486}%
\special{pa 1496 484}%
\special{pa 1500 482}%
\special{pa 1506 480}%
\special{pa 1510 480}%
\special{pa 1516 482}%
\special{pa 1520 486}%
\special{pa 1526 490}%
\special{pa 1530 494}%
\special{pa 1536 498}%
\special{pa 1540 504}%
\special{pa 1546 508}%
\special{pa 1550 512}%
\special{pa 1556 516}%
\special{pa 1560 518}%
\special{pa 1566 520}%
\special{pa 1570 520}%
\special{pa 1576 520}%
\special{pa 1580 518}%
\special{pa 1586 514}%
\special{pa 1590 510}%
\special{pa 1596 506}%
\special{ip}%
\special{pa 1600 500}%
\special{pa 1606 496}%
\special{pa 1610 490}%
\special{pa 1616 486}%
\special{pa 1620 484}%
\special{pa 1626 482}%
\special{pa 1630 480}%
\special{pa 1636 480}%
\special{pa 1640 482}%
\special{pa 1646 484}%
\special{pa 1650 488}%
\special{pa 1656 492}%
\special{pa 1660 498}%
\special{pa 1666 502}%
\special{pa 1670 508}%
\special{pa 1676 512}%
\special{pa 1680 516}%
\special{pa 1686 518}%
\special{pa 1690 520}%
\special{pa 1696 520}%
\special{pa 1700 520}%
\special{pa 1706 518}%
\special{pa 1710 514}%
\special{pa 1716 510}%
\special{pa 1720 506}%
\special{pa 1726 502}%
\special{pa 1730 496}%
\special{pa 1736 492}%
\special{pa 1740 488}%
\special{pa 1746 484}%
\special{pa 1750 482}%
\special{pa 1756 480}%
\special{pa 1760 480}%
\special{pa 1766 482}%
\special{pa 1770 484}%
\special{pa 1776 488}%
\special{pa 1780 492}%
\special{pa 1786 498}%
\special{pa 1790 502}%
\special{pa 1796 506}%
\special{pa 1800 512}%
\special{pa 1806 516}%
\special{pa 1810 518}%
\special{pa 1816 520}%
\special{pa 1820 520}%
\special{pa 1826 520}%
\special{pa 1830 518}%
\special{pa 1836 516}%
\special{pa 1840 512}%
\special{pa 1846 506}%
\special{pa 1850 502}%
\special{pa 1856 496}%
\special{pa 1860 492}%
\special{pa 1866 488}%
\special{pa 1870 484}%
\special{pa 1876 482}%
\special{pa 1880 480}%
\special{pa 1886 480}%
\special{pa 1890 482}%
\special{pa 1896 484}%
\special{pa 1900 488}%
\special{sp}%
% VECTOR 1 0 3 0
% 8 1400 200 1600 200 1800 800 1600 800 1350 600 1450 600 1460 400 1370 400
% 
\special{pn 13}%
\special{pa 1400 200}%
\special{pa 1600 200}%
\special{fp}%
\special{sh 1}%
\special{pa 1600 200}%
\special{pa 1534 180}%
\special{pa 1548 200}%
\special{pa 1534 220}%
\special{pa 1600 200}%
\special{fp}%
\special{pa 1800 800}%
\special{pa 1600 800}%
\special{fp}%
\special{sh 1}%
\special{pa 1600 800}%
\special{pa 1668 820}%
\special{pa 1654 800}%
\special{pa 1668 780}%
\special{pa 1600 800}%
\special{fp}%
\special{pa 1350 600}%
\special{pa 1450 600}%
\special{fp}%
\special{sh 1}%
\special{pa 1450 600}%
\special{pa 1384 580}%
\special{pa 1398 600}%
\special{pa 1384 620}%
\special{pa 1450 600}%
\special{fp}%
\special{pa 1460 400}%
\special{pa 1370 400}%
\special{fp}%
\special{sh 1}%
\special{pa 1370 400}%
\special{pa 1438 420}%
\special{pa 1424 400}%
\special{pa 1438 380}%
\special{pa 1370 400}%
\special{fp}%
% STR 2 0 3 0
% 3 240 440 240 540 2 0
% $qq\bar{q}\bar{q}$
\put(2.4000,-5.4000){\makebox(0,0)[lb]{$qq\bar{q}\bar{q}$}}%
% STR 2 0 3 0
% 3 2690 440 2690 540 2 0
% $q\bar{q}$
\put(26.9000,-5.4000){\makebox(0,0)[lb]{$q\bar{q}$}}%
\end{picture}%
\hspace{2cm} \\
Fig.1\ \ OZI rule allowed graph for $qq\bar{q}\bar{q}$ and $q\bar{q}$ states 
transition
\end{center}
\end{figure}
%%%%%%%%%%%%%%%%%%%%%%%%%%%%%%%%%%%
This transition is represented as 
\begin{eqnarray}
L_{\rm int}&=&-\lambda_{01}\epsilon^{abc}\epsilon_{def}N^d_a
    N'^e_b\delta^f_c=\lambda_{01}[a_0^+{a'}_0^-+a_0^-{a'}_0^++a_0^0{a'}_0^0
    +\kappa^+K_0^{*-}      \nonumber\\
    &&+\kappa^-K_0^{*+}+\kappa^0K_0^{*0}+\bar{\kappa}^+
    \bar{K}_0^{*-}-\sqrt{2}f_Nf'_N-f_Sf'_N-\sqrt{2}f_Nf'_S].
\end{eqnarray}
The inter-mixing parameter $\lambda_{01}$ represents the strength of the 
inter-mixing and can be considered as rather large. 
\par
We first consider the $I=1$ $a_0(980)$ and $a_0(1450)$ mixing. Representing 
the before mixing $qq\bar{q}\bar{q}$ state by $\overline{a_0(980)}$ and $q\bar{q}$ 
by $\overline{a_0(1450)}$ and mixing angle as $\theta_a$, physical after mixing 
state $a_0(980)$ and $a_0(1450)$ are written as follows;
\begin{equation}
\begin{array}{l}
a_0(980)=\cos\theta_a\overline{a_0(980)}-\sin\theta_a\overline{a_0(1450)}\\
a_0(1450)=\sin\theta_a\overline{a_0(980)}+\cos\theta_a\overline{a_0(1450)}
\end{array}
\end{equation} 
and the mixing matrix is represented as
\begin{equation}
\left(\begin{array}{cc}
    m^2_{\overline{a_0(980)}} & \lambda_{01}^a\\
    \lambda_{01}^a & m^2_{\overline{a_0(1450)}}
    \end{array}\right),
\end{equation}
where $m_{\overline{a_0(980)}}$ and $m_{\overline{a_0(1450)}}$ are the 
before mixing masses for $a_0(980)$ and $a_0(1450)$ states. 
For the values of $m_{\overline{a_0(980)}}$ and $m_{\overline{a_0(1450)}}$, 
in the first our work \cite{teshima1}, we adopted the values 
$$ m_{\overline{a_0(980)}}=1271\pm31 {\rm MeV},\ \ 
   m_{\overline{a_0(1450)}}=1236\pm20 {\rm MeV}, $$ 
estimated from the relation 
$m^2(2^{++})-m^2(1^{++})=2(m^2(1^{++})-m^2(0^{++}))$   
resulting from the $L\cdot S$ force.
Diagonalising the mass matrix in Eq. (4) and taking the eigenvalues of masses 
\begin{equation}
m_{a_0(980)}=984.8\pm1.4{\rm MeV},\ \ m_{a_0(1450)}=1474\pm19{\rm MeV},
\end{equation}
we can get the result
\begin{equation}
\lambda_{01}^a=0.600\pm0.028{\rm GeV^2}, \ \  \ \ 
\theta_{a}=47.1\pm3.5^\circ. 
\end{equation}
Similarly, we estimated the strength $\lambda^K_{01}$ and mixing angle $\theta_K$ 
for $I=1/2$ $\kappa(900)$ and $K^*_0(1430)$ mixing case. 
Using the mass values before mixing and after mixing, 
\begin{equation}
\begin{array}{l}
m_{\overline{\kappa(900)}}=1047\pm62{\rm MeV},\ \ m_{\overline{K^*_0(1430)}}=
1307\pm11{\rm MeV}, \\
m_{\kappa(900)}=900\pm70{\rm MeV},\ \ m_{K^*_0(1430)}=1412\pm6{\rm MeV}, 
\end{array}
\end{equation}
we get the results
\begin{equation}
\lambda_{01}^K=0.507\pm84{\rm GeV^2},\ \ \ \  
\theta_{K}=29.5\pm15.5^\circ.
\end{equation}
\par 
In our next work \cite{teshima2}, we estimated the mixing angle $\theta_a$ and 
$\theta_K$ analyzing the $a_0(980)$, $a_0(1450)$ and $K^*_0(1430)$ decay processes 
to two pseudoscalar meson, and get 
the results, 
\begin{equation}
\theta_a=\theta_K=(9\pm4)^{\circ}.
\end{equation} 
The value of $\lambda_{01}$ and mass values before mixing of $a_0(980)$, 
$a_0(1450)$ and $\kappa(900)$, $K_0^*(1430)$ for this mixing angle are estimated 
as follows; 
\begin{eqnarray}
&&\lambda_{01}^a=\lambda_{01}^K=0.19\makebox{$ {+0.07\atop-0.09}$} {\rm GeV^2},
\nonumber 
\\
&&m_{\overline{a_0(980)}}=1.00\makebox{$ {+0.02\atop-0.01}$}{\rm GeV},\ \ 
m_{\overline{a_0(1450)}}=1.46\pm0.01{\rm GeV}\nonumber,
\\
&&m_{\overline{K_0(900)}}=0.92\pm0.01{\rm GeV},\ \ 
m_{\overline{K^*_0(1430)}}=1.40\pm0.01{\rm GeV}.
\end{eqnarray} 
%%%%%%%%%%%%%%%%(2-2)%%%%%%%%%%%%%%%%%%%%%%
\subsection{Inter-mixing between $I=0$ low and high mass scalar mesons}
Among the $I=0, L=1\ q\bar{q}$ scalar mesons, there are the intra-mixing weaker 
than the inter-mixing, caused from the transition between themselves represented by 
the OZI rule suppression graph shown in Fig. 2, and furthermore the mixing between 
the $q\bar{q}$ scalar meson and the glueball caused from the transition represented 
by the graph shown in Fig. 3. Thus, the mass matrix for these $I=0, L=1\ q\bar{q}$ 
scalar mesons and glueball is represented as 
\begin{equation}
\left(\begin{array}{ccc}
m^2_{N'}+2\lambda_1&\sqrt{2}\lambda_1&\sqrt{2}\lambda_G\\
\sqrt{2}\lambda_1&2m^2_{S'}+\lambda_1&\lambda_G\\ 
\sqrt{2}\lambda_G&\lambda_G&\lambda_{GG}
\end{array}\right), 
\end{equation}
where $m^2_{N'}=m^2_{\overline{a'_0}}$, $m^2_{S'}=2m^2_{\overline{K'_0}}-
m^2_{\overline{a'_0}}$, and $\lambda_1$ is the transition strength among the 
$I=0,\ q\bar{q}$ mesons, $\lambda_{G}$ is the transition strength between 
$q\bar{q}$ and glueball $gg$ and $\lambda_{GG}$ is the pure glueball mass square.
%%%%%%%%%%%%%%%%%%%%%%%%%%%%%%%%%%
\begin{figure}
\vspace{0.5cm}
\begin{center}
%WinTpicVersion3.08
\unitlength 0.1in
\begin{picture}( 26.0000, 11.0300)(  3.0000,-12.3300)
% FUNC 1 0 3 0
% 9 1419 170 1955 650 1419 410 1686 410 1419 170 1419 170 1955 650 0 3 0 0
% 0.1sin(10x)
\special{pn 13}%
\special{pa 1416 414}%
\special{pa 1420 410}%
\special{pa 1426 406}%
\special{pa 1430 400}%
\special{pa 1436 396}%
\special{pa 1440 394}%
\special{pa 1446 390}%
\special{pa 1450 388}%
\special{pa 1456 388}%
\special{pa 1460 386}%
\special{pa 1466 386}%
\special{pa 1470 388}%
\special{pa 1476 390}%
\special{pa 1480 392}%
\special{pa 1486 396}%
\special{pa 1490 400}%
\special{pa 1496 404}%
\special{pa 1500 408}%
\special{pa 1506 412}%
\special{pa 1510 416}%
\special{pa 1516 422}%
\special{pa 1520 424}%
\special{pa 1526 428}%
\special{pa 1530 430}%
\special{pa 1536 432}%
\special{pa 1540 434}%
\special{pa 1546 434}%
\special{pa 1550 434}%
\special{pa 1556 432}%
\special{pa 1560 430}%
\special{pa 1566 428}%
\special{pa 1570 424}%
\special{pa 1576 420}%
\special{pa 1580 416}%
\special{pa 1586 412}%
\special{pa 1590 408}%
\special{pa 1596 404}%
\special{pa 1600 400}%
\special{pa 1606 396}%
\special{pa 1610 392}%
\special{pa 1616 390}%
\special{pa 1620 388}%
\special{pa 1626 386}%
\special{pa 1630 386}%
\special{pa 1636 388}%
\special{pa 1640 388}%
\special{pa 1646 390}%
\special{pa 1650 394}%
\special{pa 1656 398}%
\special{pa 1660 402}%
\special{pa 1666 406}%
\special{pa 1670 410}%
\special{pa 1676 414}%
\special{pa 1680 418}%
\special{pa 1686 422}%
\special{pa 1690 426}%
\special{pa 1696 430}%
\special{pa 1700 432}%
\special{pa 1706 434}%
\special{pa 1710 434}%
\special{pa 1716 434}%
\special{pa 1720 434}%
\special{pa 1726 432}%
\special{pa 1730 430}%
\special{pa 1736 426}%
\special{pa 1740 422}%
\special{pa 1746 418}%
\special{pa 1750 414}%
\special{pa 1756 410}%
\special{pa 1760 406}%
\special{pa 1766 402}%
\special{pa 1770 398}%
\special{pa 1776 394}%
\special{pa 1780 390}%
\special{pa 1786 388}%
\special{pa 1790 388}%
\special{pa 1796 386}%
\special{pa 1800 386}%
\special{pa 1806 388}%
\special{pa 1810 390}%
\special{pa 1816 392}%
\special{pa 1820 396}%
\special{pa 1826 398}%
\special{pa 1830 404}%
\special{pa 1836 408}%
\special{pa 1840 412}%
\special{pa 1846 416}%
\special{pa 1850 420}%
\special{pa 1856 424}%
\special{pa 1860 428}%
\special{pa 1866 430}%
\special{pa 1870 432}%
\special{pa 1876 434}%
\special{pa 1880 434}%
\special{pa 1886 434}%
\special{pa 1890 432}%
\special{pa 1896 430}%
\special{pa 1900 428}%
\special{pa 1906 424}%
\special{pa 1910 422}%
\special{pa 1916 416}%
\special{pa 1920 412}%
\special{pa 1926 408}%
\special{pa 1930 404}%
\special{pa 1936 400}%
\special{pa 1940 396}%
\special{pa 1946 392}%
\special{pa 1950 390}%
\special{pa 1956 388}%
\special{sp}%
% LINE 1 0 3 0
% 16 620 395 1426 395 1426 996 1426 996 620 996 620 996 620 996 1411 996 1411 996 1411 395 1948 395 2754 395 2754 996 1948 996 1948 996 1948 395
% 
\special{pn 13}%
\special{pa 620 396}%
\special{pa 1426 396}%
\special{fp}%
\special{pa 1426 996}%
\special{pa 1426 996}%
\special{fp}%
\special{pa 620 996}%
\special{pa 620 996}%
\special{fp}%
\special{pa 620 996}%
\special{pa 1412 996}%
\special{fp}%
\special{pa 1412 996}%
\special{pa 1412 396}%
\special{fp}%
\special{pa 1948 396}%
\special{pa 2754 396}%
\special{fp}%
\special{pa 2754 996}%
\special{pa 1948 996}%
\special{fp}%
\special{pa 1948 996}%
\special{pa 1948 396}%
\special{fp}%
% FUNC 1 0 3 0
% 9 1407 753 1944 1233 1422 994 1689 994 1422 753 1407 753 1944 1233 0 3 0 0
% 0.1sin(10x)
\special{pn 13}%
\special{pa 1406 1008}%
\special{pa 1410 1004}%
\special{pa 1416 1000}%
\special{pa 1420 996}%
\special{pa 1426 992}%
\special{pa 1430 988}%
\special{pa 1436 984}%
\special{pa 1440 980}%
\special{pa 1446 976}%
\special{pa 1450 974}%
\special{pa 1456 972}%
\special{pa 1460 970}%
\special{pa 1466 970}%
\special{pa 1470 972}%
\special{pa 1476 972}%
\special{pa 1480 974}%
\special{pa 1486 978}%
\special{pa 1490 980}%
\special{pa 1496 984}%
\special{pa 1500 990}%
\special{pa 1506 994}%
\special{pa 1510 998}%
\special{pa 1516 1002}%
\special{pa 1520 1006}%
\special{pa 1526 1010}%
\special{pa 1530 1014}%
\special{pa 1536 1016}%
\special{pa 1540 1018}%
\special{pa 1546 1018}%
\special{pa 1550 1018}%
\special{pa 1556 1018}%
\special{pa 1560 1016}%
\special{pa 1566 1014}%
\special{pa 1570 1010}%
\special{pa 1576 1008}%
\special{pa 1580 1004}%
\special{pa 1586 998}%
\special{pa 1590 994}%
\special{pa 1596 990}%
\special{pa 1600 986}%
\special{pa 1606 982}%
\special{pa 1610 978}%
\special{pa 1616 974}%
\special{pa 1620 972}%
\special{pa 1626 972}%
\special{pa 1630 970}%
\special{pa 1636 970}%
\special{pa 1640 972}%
\special{pa 1646 974}%
\special{pa 1650 976}%
\special{pa 1656 980}%
\special{pa 1660 982}%
\special{pa 1666 986}%
\special{pa 1670 992}%
\special{pa 1676 996}%
\special{pa 1680 1000}%
\special{pa 1686 1004}%
\special{pa 1690 1008}%
\special{pa 1696 1012}%
\special{pa 1700 1014}%
\special{pa 1706 1016}%
\special{pa 1710 1018}%
\special{pa 1716 1018}%
\special{pa 1720 1018}%
\special{pa 1726 1018}%
\special{pa 1730 1016}%
\special{pa 1736 1012}%
\special{pa 1740 1010}%
\special{pa 1746 1006}%
\special{pa 1750 1002}%
\special{pa 1756 996}%
\special{pa 1760 992}%
\special{pa 1766 988}%
\special{pa 1770 984}%
\special{pa 1776 980}%
\special{pa 1780 976}%
\special{pa 1786 974}%
\special{pa 1790 972}%
\special{pa 1796 970}%
\special{pa 1800 970}%
\special{pa 1806 970}%
\special{pa 1810 972}%
\special{pa 1816 974}%
\special{pa 1820 978}%
\special{pa 1826 980}%
\special{pa 1830 984}%
\special{pa 1836 988}%
\special{pa 1840 994}%
\special{pa 1846 998}%
\special{pa 1850 1002}%
\special{pa 1856 1006}%
\special{pa 1860 1010}%
\special{pa 1866 1014}%
\special{pa 1870 1016}%
\special{pa 1876 1018}%
\special{pa 1880 1018}%
\special{pa 1886 1018}%
\special{pa 1890 1018}%
\special{pa 1896 1016}%
\special{pa 1900 1014}%
\special{pa 1906 1012}%
\special{pa 1910 1008}%
\special{pa 1916 1004}%
\special{pa 1920 1000}%
\special{pa 1926 994}%
\special{pa 1930 990}%
\special{pa 1936 986}%
\special{pa 1940 982}%
\special{sp}%
% BOX 2 5 3 0
% 2 1136 410 600 1011
% 
\special{pn 8}%
\special{pa 1136 410}%
\special{pa 600 410}%
\special{pa 600 1012}%
\special{pa 1136 1012}%
\special{pa 1136 410}%
\special{ip}%
% BOX 2 5 3 0
% 2 2224 410 2760 1011
% 
\special{pn 8}%
\special{pa 2224 410}%
\special{pa 2760 410}%
\special{pa 2760 1012}%
\special{pa 2224 1012}%
\special{pa 2224 410}%
\special{ip}%
% LINE 2 0 3 0
% 14 2760 457 2224 939 2760 603 2316 998 2760 746 2478 998 2760 889 2639 998 2653 410 2224 795 2490 410 2224 650 2331 410 2224 506
% 
\special{pn 8}%
\special{pa 2760 458}%
\special{pa 2224 940}%
\special{fp}%
\special{pa 2760 604}%
\special{pa 2316 998}%
\special{fp}%
\special{pa 2760 746}%
\special{pa 2478 998}%
\special{fp}%
\special{pa 2760 890}%
\special{pa 2640 998}%
\special{fp}%
\special{pa 2654 410}%
\special{pa 2224 796}%
\special{fp}%
\special{pa 2490 410}%
\special{pa 2224 650}%
\special{fp}%
\special{pa 2332 410}%
\special{pa 2224 506}%
\special{fp}%
% LINE 3 0 3 0
% 14 1136 614 709 998 1136 471 600 949 1043 410 600 806 882 410 600 662 721 410 600 517 1136 758 868 998 1136 903 1029 998
% 
\special{pn 4}%
\special{pa 1136 614}%
\special{pa 710 998}%
\special{fp}%
\special{pa 1136 472}%
\special{pa 600 950}%
\special{fp}%
\special{pa 1044 410}%
\special{pa 600 806}%
\special{fp}%
\special{pa 882 410}%
\special{pa 600 662}%
\special{fp}%
\special{pa 722 410}%
\special{pa 600 518}%
\special{fp}%
\special{pa 1136 758}%
\special{pa 868 998}%
\special{fp}%
\special{pa 1136 904}%
\special{pa 1030 998}%
\special{fp}%
% LINE 2 5 3 0
% 2 2236 410 2236 1011
% 
\special{pn 8}%
\special{pa 2236 410}%
\special{pa 2236 1012}%
\special{ip}%
% LINE 2 0 3 0
% 4 2214 394 2214 995 1140 995 1140 394
% 
\special{pn 8}%
\special{pa 2214 394}%
\special{pa 2214 996}%
\special{fp}%
\special{pa 1140 996}%
\special{pa 1140 394}%
\special{fp}%
% STR 2 0 3 0
% 3 300 700 300 800 2 0
% $q\bar{q}$
\put(3.0000,-8.0000){\makebox(0,0)[lb]{$q\bar{q}$}}%
% STR 2 0 3 0
% 3 2900 700 2900 800 2 0
% $q\bar{q}$
\put(29.0000,-8.0000){\makebox(0,0)[lb]{$q\bar{q}$}}%
% STR 2 0 3 0
% 3 1600 200 1600 300 2 0
% $g$
\put(16.0000,-3.0000){\makebox(0,0)[lb]{$g$}}%
% STR 2 0 3 0
% 3 1600 1150 1600 1250 2 0
% $g$
\put(16.0000,-12.5000){\makebox(0,0)[lb]{$g$}}%
% VECTOR 1 0 3 0
% 2 990 400 1000 400
% 
\special{pn 13}%
\special{pa 990 400}%
\special{pa 1000 400}%
\special{fp}%
\special{sh 1}%
\special{pa 1000 400}%
\special{pa 934 380}%
\special{pa 948 400}%
\special{pa 934 420}%
\special{pa 1000 400}%
\special{fp}%
% VECTOR 1 0 3 0
% 2 930 1000 920 1000
% 
\special{pn 13}%
\special{pa 930 1000}%
\special{pa 920 1000}%
\special{fp}%
\special{sh 1}%
\special{pa 920 1000}%
\special{pa 988 1020}%
\special{pa 974 1000}%
\special{pa 988 980}%
\special{pa 920 1000}%
\special{fp}%
% VECTOR 1 0 3 0
% 2 2390 400 2400 400
% 
\special{pn 13}%
\special{pa 2390 400}%
\special{pa 2400 400}%
\special{fp}%
\special{sh 1}%
\special{pa 2400 400}%
\special{pa 2334 380}%
\special{pa 2348 400}%
\special{pa 2334 420}%
\special{pa 2400 400}%
\special{fp}%
% VECTOR 1 0 3 0
% 2 2330 1000 2320 1000
% 
\special{pn 13}%
\special{pa 2330 1000}%
\special{pa 2320 1000}%
\special{fp}%
\special{sh 1}%
\special{pa 2320 1000}%
\special{pa 2388 1020}%
\special{pa 2374 1000}%
\special{pa 2388 980}%
\special{pa 2320 1000}%
\special{fp}%
\end{picture}%
\\
Fig. 2. OZI rule suppression graph for $q\bar{q}-q\bar{q}$ transition.
\end{center}
\end{figure}
%%%%%%%%%%%%%%%%%%%%%%%%%%%%%%%%%%
%%%%%%%%%%%%%%%%%%%%%%%%%%%%%%%%%%
\begin{figure}
\begin{center}
\input{fig3}
\vspace{0.5cm}\\
Fig. 3. Transition graph between (a) $q\bar{q}$ and $gg$, and (b) $gg$ and $gg$.
\end{center}
\end{figure}
%%%%%%%%%%%%%%%%%%%%%%%%%%%%%%%%%%
\par
For the light $I=0\ qq\bar{q}\bar{q}$ scalar mesons, there are the intra-mixing 
caused from the transition between themselves represented by the OZI rule 
suppression graph shown in Fig. 4, and the mass matrix for these $I=0\ qq\bar{q}
\bar{q}$ scalar meson is represented as  
\begin{equation}
\left(\begin{array}{cc}
m^2_{NN}+2\lambda_0&\sqrt{2}\lambda_0\\
\sqrt{2}\lambda_0&2m^2_{NS}+\lambda_0 
\end{array}\right), 
\end{equation}
where $m^2_{NN}=2m^2_{\overline{K_0}}-m^2_{\overline{a_0}}$, $m^2_{NS}=
m^2_{\overline{a_0}}$, 
and $\lambda_0$ represents the transition strength between $I=0\ \ qq\bar{q}
\bar{q}$ mesons.  
%%%%%%%%%%%%%%%%%%%%%%%%%%%%%%%%%%
\begin{figure}
\begin{center}
%WinTpicVersion3.08
\unitlength 0.1in
\begin{picture}( 25.0000,  6.1000)(  2.0000,-10.1000)
% LINE 1 0 3 0
% 12 600 410 2600 410 2600 1010 600 1010 600 610 1300 610 1300 810 600 810 2600 610 1900 610 1900 810 2600 810
% 
\special{pn 13}%
\special{pa 600 410}%
\special{pa 2600 410}%
\special{fp}%
\special{pa 2600 1010}%
\special{pa 600 1010}%
\special{fp}%
\special{pa 600 610}%
\special{pa 1300 610}%
\special{fp}%
\special{pa 1300 810}%
\special{pa 600 810}%
\special{fp}%
\special{pa 2600 610}%
\special{pa 1900 610}%
\special{fp}%
\special{pa 1900 810}%
\special{pa 2600 810}%
\special{fp}%
% CIRCLE 1 0 3 0
% 4 1900 710 1900 610 1900 610 1900 810
% 
\special{pn 13}%
\special{ar 1900 710 100 100  1.5707963 4.7123890}%
% CIRCLE 1 0 3 0
% 4 1300 710 1300 810 1300 810 1300 610
% 
\special{pn 13}%
\special{ar 1300 710 100 100  4.7123890 6.2831853}%
\special{ar 1300 710 100 100  0.0000000 1.5707963}%
% FUNC 1 0 3 0
% 9 1400 610 1800 810 1600 710 1800 710 1600 510 1400 610 1800 810 0 3 0 0
% 0.1sin(10x)
\special{pn 13}%
\special{pa 1400 700}%
\special{pa 1406 704}%
\special{pa 1410 708}%
\special{pa 1416 714}%
\special{pa 1420 718}%
\special{pa 1426 722}%
\special{pa 1430 726}%
\special{pa 1436 728}%
\special{pa 1440 730}%
\special{pa 1446 730}%
\special{pa 1450 730}%
\special{pa 1456 726}%
\special{pa 1460 724}%
\special{pa 1466 720}%
\special{pa 1470 714}%
\special{pa 1476 710}%
\special{pa 1480 704}%
\special{pa 1486 700}%
\special{pa 1490 696}%
\special{pa 1496 694}%
\special{pa 1500 692}%
\special{pa 1506 690}%
\special{pa 1510 690}%
\special{pa 1516 692}%
\special{pa 1520 696}%
\special{pa 1526 700}%
\special{pa 1530 704}%
\special{pa 1536 708}%
\special{pa 1540 714}%
\special{pa 1546 718}%
\special{pa 1550 722}%
\special{pa 1556 726}%
\special{pa 1560 728}%
\special{pa 1566 730}%
\special{pa 1570 730}%
\special{pa 1576 730}%
\special{pa 1580 728}%
\special{pa 1586 724}%
\special{pa 1590 720}%
\special{pa 1596 716}%
\special{pa 1600 710}%
\special{pa 1606 706}%
\special{pa 1610 700}%
\special{pa 1616 696}%
\special{pa 1620 694}%
\special{pa 1626 692}%
\special{pa 1630 690}%
\special{pa 1636 690}%
\special{pa 1640 692}%
\special{pa 1646 694}%
\special{pa 1650 698}%
\special{pa 1656 702}%
\special{pa 1660 708}%
\special{pa 1666 712}%
\special{pa 1670 718}%
\special{pa 1676 722}%
\special{pa 1680 726}%
\special{pa 1686 728}%
\special{pa 1690 730}%
\special{pa 1696 730}%
\special{pa 1700 730}%
\special{pa 1706 728}%
\special{pa 1710 724}%
\special{pa 1716 720}%
\special{pa 1720 716}%
\special{pa 1726 712}%
\special{pa 1730 706}%
\special{pa 1736 702}%
\special{pa 1740 698}%
\special{pa 1746 694}%
\special{pa 1750 692}%
\special{pa 1756 690}%
\special{pa 1760 690}%
\special{pa 1766 692}%
\special{pa 1770 694}%
\special{pa 1776 698}%
\special{pa 1780 702}%
\special{pa 1786 708}%
\special{pa 1790 712}%
\special{pa 1796 716}%
\special{pa 1800 722}%
\special{sp}%
% BOX 1 5 3 0
% 2 600 410 1200 1010
% 
\special{pn 13}%
\special{pa 600 410}%
\special{pa 1200 410}%
\special{pa 1200 1010}%
\special{pa 600 1010}%
\special{pa 600 410}%
\special{ip}%
% BOX 1 5 3 0
% 2 590 410 1190 1010
% 
\special{pn 13}%
\special{pa 590 410}%
\special{pa 1190 410}%
\special{pa 1190 1010}%
\special{pa 590 1010}%
\special{pa 590 410}%
\special{ip}%
% LINE 3 0 3 0
% 12 1120 810 930 1000 1000 810 810 1000 880 810 690 1000 760 810 600 970 640 810 600 850 1180 870 1050 1000
% 
\special{pn 4}%
\special{pa 1120 810}%
\special{pa 930 1000}%
\special{fp}%
\special{pa 1000 810}%
\special{pa 810 1000}%
\special{fp}%
\special{pa 880 810}%
\special{pa 690 1000}%
\special{fp}%
\special{pa 760 810}%
\special{pa 600 970}%
\special{fp}%
\special{pa 640 810}%
\special{pa 600 850}%
\special{fp}%
\special{pa 1180 870}%
\special{pa 1050 1000}%
\special{fp}%
% LINE 3 0 3 0
% 12 1080 610 890 800 960 610 770 800 840 610 650 800 720 610 600 730 1180 630 1010 800 1180 750 1130 800
% 
\special{pn 4}%
\special{pa 1080 610}%
\special{pa 890 800}%
\special{fp}%
\special{pa 960 610}%
\special{pa 770 800}%
\special{fp}%
\special{pa 840 610}%
\special{pa 650 800}%
\special{fp}%
\special{pa 720 610}%
\special{pa 600 730}%
\special{fp}%
\special{pa 1180 630}%
\special{pa 1010 800}%
\special{fp}%
\special{pa 1180 750}%
\special{pa 1130 800}%
\special{fp}%
% LINE 3 0 3 0
% 12 1040 410 850 600 920 410 730 600 800 410 610 600 680 410 600 490 1160 410 970 600 1180 510 1090 600
% 
\special{pn 4}%
\special{pa 1040 410}%
\special{pa 850 600}%
\special{fp}%
\special{pa 920 410}%
\special{pa 730 600}%
\special{fp}%
\special{pa 800 410}%
\special{pa 610 600}%
\special{fp}%
\special{pa 680 410}%
\special{pa 600 490}%
\special{fp}%
\special{pa 1160 410}%
\special{pa 970 600}%
\special{fp}%
\special{pa 1180 510}%
\special{pa 1090 600}%
\special{fp}%
% LINE 2 0 3 0
% 4 1200 410 1200 1010 2000 1010 2000 410
% 
\special{pn 8}%
\special{pa 1200 410}%
\special{pa 1200 1010}%
\special{fp}%
\special{pa 2000 1010}%
\special{pa 2000 410}%
\special{fp}%
% BOX 2 5 3 0
% 2 2000 410 2600 1010
% 
\special{pn 8}%
\special{pa 2000 410}%
\special{pa 2600 410}%
\special{pa 2600 1010}%
\special{pa 2000 1010}%
\special{pa 2000 410}%
\special{ip}%
% BOX 2 5 3 0
% 2 2000 410 2600 1010
% 
\special{pn 8}%
\special{pa 2000 410}%
\special{pa 2600 410}%
\special{pa 2600 1010}%
\special{pa 2000 1010}%
\special{pa 2000 410}%
\special{ip}%
% LINE 3 0 3 0
% 12 2440 810 2250 1000 2320 810 2130 1000 2200 810 2010 1000 2080 810 2000 890 2560 810 2370 1000 2600 890 2490 1000
% 
\special{pn 4}%
\special{pa 2440 810}%
\special{pa 2250 1000}%
\special{fp}%
\special{pa 2320 810}%
\special{pa 2130 1000}%
\special{fp}%
\special{pa 2200 810}%
\special{pa 2010 1000}%
\special{fp}%
\special{pa 2080 810}%
\special{pa 2000 890}%
\special{fp}%
\special{pa 2560 810}%
\special{pa 2370 1000}%
\special{fp}%
\special{pa 2600 890}%
\special{pa 2490 1000}%
\special{fp}%
% BOX 2 5 3 0
% 2 2000 400 2600 800
% 
\special{pn 8}%
\special{pa 2000 400}%
\special{pa 2600 400}%
\special{pa 2600 800}%
\special{pa 2000 800}%
\special{pa 2000 400}%
\special{ip}%
% LINE 3 0 3 0
% 12 2280 610 2090 800 2160 610 2000 770 2040 610 2000 650 2400 610 2210 800 2520 610 2330 800 2600 650 2450 800
% 
\special{pn 4}%
\special{pa 2280 610}%
\special{pa 2090 800}%
\special{fp}%
\special{pa 2160 610}%
\special{pa 2000 770}%
\special{fp}%
\special{pa 2040 610}%
\special{pa 2000 650}%
\special{fp}%
\special{pa 2400 610}%
\special{pa 2210 800}%
\special{fp}%
\special{pa 2520 610}%
\special{pa 2330 800}%
\special{fp}%
\special{pa 2600 650}%
\special{pa 2450 800}%
\special{fp}%
% LINE 3 0 3 0
% 12 2360 410 2170 600 2240 410 2050 600 2120 410 2000 530 2480 410 2290 600 2590 420 2410 600 2600 530 2530 600
% 
\special{pn 4}%
\special{pa 2360 410}%
\special{pa 2170 600}%
\special{fp}%
\special{pa 2240 410}%
\special{pa 2050 600}%
\special{fp}%
\special{pa 2120 410}%
\special{pa 2000 530}%
\special{fp}%
\special{pa 2480 410}%
\special{pa 2290 600}%
\special{fp}%
\special{pa 2590 420}%
\special{pa 2410 600}%
\special{fp}%
\special{pa 2600 530}%
\special{pa 2530 600}%
\special{fp}%
% VECTOR 1 0 3 0
% 8 1200 610 1300 610 1300 810 1200 810 1880 610 1980 610 1980 810 1890 810
% 
\special{pn 13}%
\special{pa 1200 610}%
\special{pa 1300 610}%
\special{fp}%
\special{sh 1}%
\special{pa 1300 610}%
\special{pa 1234 590}%
\special{pa 1248 610}%
\special{pa 1234 630}%
\special{pa 1300 610}%
\special{fp}%
\special{pa 1300 810}%
\special{pa 1200 810}%
\special{fp}%
\special{sh 1}%
\special{pa 1200 810}%
\special{pa 1268 830}%
\special{pa 1254 810}%
\special{pa 1268 790}%
\special{pa 1200 810}%
\special{fp}%
\special{pa 1880 610}%
\special{pa 1980 610}%
\special{fp}%
\special{sh 1}%
\special{pa 1980 610}%
\special{pa 1914 590}%
\special{pa 1928 610}%
\special{pa 1914 630}%
\special{pa 1980 610}%
\special{fp}%
\special{pa 1980 810}%
\special{pa 1890 810}%
\special{fp}%
\special{sh 1}%
\special{pa 1890 810}%
\special{pa 1958 830}%
\special{pa 1944 810}%
\special{pa 1958 790}%
\special{pa 1890 810}%
\special{fp}%
% STR 2 0 3 0
% 3 2700 660 2700 760 2 0
% $qq\bar{q}\bar{q}$
\put(27.0000,-7.6000){\makebox(0,0)[lb]{$qq\bar{q}\bar{q}$}}%
% STR 2 0 3 0
% 3 200 660 200 760 2 0
% $qq\bar{q}\bar{q}$
\put(2.0000,-7.6000){\makebox(0,0)[lb]{$qq\bar{q}\bar{q}$}}%
% VECTOR 1 0 3 0
% 4 1800 1010 1600 1010 1400 410 1600 410
% 
\special{pn 13}%
\special{pa 1800 1010}%
\special{pa 1600 1010}%
\special{fp}%
\special{sh 1}%
\special{pa 1600 1010}%
\special{pa 1668 1030}%
\special{pa 1654 1010}%
\special{pa 1668 990}%
\special{pa 1600 1010}%
\special{fp}%
\special{pa 1400 410}%
\special{pa 1600 410}%
\special{fp}%
\special{sh 1}%
\special{pa 1600 410}%
\special{pa 1534 390}%
\special{pa 1548 410}%
\special{pa 1534 430}%
\special{pa 1600 410}%
\special{fp}%
\end{picture}%
\vspace{0.5cm}\\
Fig. 4. OZI suppression graph for $qq\bar{q}\bar{q}-qq\bar{q}\bar{q}$ transition.
\end{center}
\end{figure}
%%%%%%%%%%%%%%%%%%%%%%%%%%%%%%%%%%
\par
The inter- and intra-mixing among $I=0$ low mass and high mass scalar mesons and 
glueball is expressed by the overall mixing mass matrix as
\begin{equation}
\left(\begin{array}{ccccc}
    m^2_{NN}+\lambda_0&\sqrt{2}\lambda_0&\sqrt{2}\lambda_{01}&0&0\\
    \sqrt{2}\lambda_0&m^2_{NS}+2\lambda_0&\lambda_{01}&\sqrt{2}\lambda_{01}&0\\
    \sqrt{2}\lambda_{01}&\lambda_{01}&m^2_{N'}+2\lambda_1&\sqrt{2}\lambda_1&
    \sqrt{2}\lambda_{G}\\
    0&\sqrt{2}\lambda_{01}&\sqrt{2}\lambda_1&m^2_{S'}+\lambda_1&
    \lambda_{G}\\
    0&0&\sqrt{2}\lambda_{G}&\lambda_{G}&\lambda_{GG}
    \end{array}\right).
\end{equation}
We use the values of $\lambda_{01}=0.19{\rm GeV^2}$ tentatively, corresponding to 
the mixing angle $\theta_{a}=\theta_{K}=9^{\circ}$ estimated in analyses of 
the decay processes as shown in Eq. (10), and then diagonalize this $5\times5$ 
mass matrix. In this case, we input the values for $m_{NN}$ etc. as follows;
\begin{equation}
    m_{NN}=0.826{\rm GeV}, \ m_{NS}=1.00{\rm GeV}, \  m_{N'}=1.46{\rm GeV}, \ 
    m_{S'}=1.34{\rm GeV}, 
\end{equation}
predicted from the relations, $m^2_{NN}=2m^2_{\overline{K_0}}-
m^2_{\overline{a_0}}$, $m^2_{NS}=m^2_{\overline{a_0}}$ and 
$m^2_{S'}=2m^2_{\overline{K'_0}}-m^2_{\overline{a'_0}}$, $m^2_{N'}=
m^2_{\overline{a'_0}}$
and the estimated values Eq. (10).
Varying the parameters $\lambda_{0}$, $\lambda_{1}$, $\lambda_{G}$ and 
$\lambda_{GG}$, we get the best fit eigenvalues for the mass of $f_0(600)$, 
$f_0(980)$, $f_0(1370)$, $f_0(1500)$ and $f_0(1710)$, 
\begin{equation}
\begin{array}{l}
     m_{f_0(600)}=0.77(0.80\pm0.40){\rm GeV},\ m_{f_0(980)}=0.93(0.980\pm0.010)
     {\rm GeV}, \\  
    m_{f_0(1370)}=1.37(1.350\pm0.150){\rm GeV},\ m_{f_0(1500)}=1.51(1.507\pm0.015)
    {\rm GeV}, \\
    m_{f_0(1710)}=1.71(1.714\pm0.005){\rm GeV}, 
\end{array}    
\end{equation}
where the values in parenthesis are quoted from \cite{PDG}. 
Best-fit values are obtained for the values of $\lambda_{0}$ etc., 
\begin{equation}
\lambda_{0}=-0.03{\rm GeV^2}, \lambda_{1}=0.04{\rm GeV^2}, \lambda_{G}=0.1
{\rm GeV^2}, \lambda_{GG}=1.7^2{\rm GeV^2}, 
\end{equation}
and at this time, mixing parameters are calculated as 
\begin{eqnarray}
&&\left(\begin{array}{c}
    f_0(600)\\
    f_0(980)\\
    f_0(1370)\\
    f_0(1500)\\
    f_0(1710)
    \end{array}\right)=[R_{f_0(M)I}]\left(\begin{array}{c}
    f_{NN}\\
    f_{NS}\\
    f_{N'}\\
    f_{S'}\\
    f_G
    \end{array}\right),\hspace{5cm}\\    
&&  [R_{f_0(M)I}]=
  \left(\begin{array}{ccccc}
     0.949&0.250&-0.185&-0.047&0.013\\
     0.270&-0.925&0.067&0.257&-0.017\\
     0.054&-0.233&0.210&-0.946& 0.064\\
     0.150&0.161&0.932&0.160&-0.239\\
     0.025&0.035&0.220&0.107&0.969     
     \end{array}\right).\nonumber
\end{eqnarray}
The mixing parameters $[R_{f_0(M)I}]$ for $f_0(600)$ and $f_0(980)$ are similar 
to those obtained in our previous works \cite{teshima1}, but the ones for 
$f_0(1370)$, $f_0(1500)$ and $f_0(1710)$ are rather different from them. This comes 
about from the smaller values of $\lambda_{01}$ in the present analysis. Because 
only the mixing parameters $[R_{f_0(M)I}]$ of $f_0(980)$ and mixing angle 
$\theta_a$ between $a_0(980)$ and $f_0(1450)$ are necessary in this work, the 
deviation of the mixing parameters $[R_{f_0(M)I}]$ for $f_0(1370)$, $f_0(1500)$ 
and $f_0(1710)$ from previous result does not get into trouble. 
%%%%%%%% section 3%%%%%%%%%%%%%%%%%%%%%%%%
\section{Radiative decays involving \mbox{$f_0(980)$} and $a_0(980)$}
\subsection{Radiative decays involving \mbox{$f_0(980)$} and $a_0(980)$ in VDM}
\noindent
In this work, we analyze the radiative decays involving \mbox{$f_0(980)$} and 
$a_0(980)$; $\phi(1020)\to f_0(980)\gamma$, $\phi(1020)\to a_0(980)\gamma$ and 
$f_0(980)\to\gamma\gamma$, $f_0(980)\to\gamma\gamma$. We assume the vector meson 
dominance model (VDM) in this analysis, for the radiative processes 
involving pseudoscalar mesons, $P\to\gamma\gamma$, $V\to P\gamma$ and $P\to 
V\gamma$ have been  interpreted well in VDM \cite{VDM}. In VDM, $V\to 
S\gamma$ and  $S\to \gamma\gamma$ processes are described by the diagram shown 
in Fig.5.
\vspace{0.0cm}
%%%%%%%%%%%%%%%%Fig.5%%%%%%%%%%%%%%%%%%%%%
\begin{center}
\input{fig5}\\
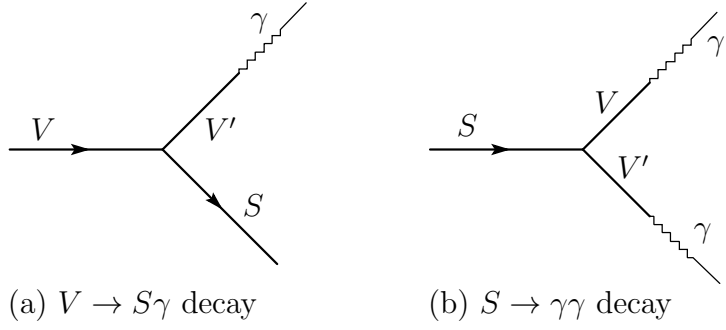
Fig. 5. Diagram for $V\to S\gamma$ and $S\to\gamma\gamma$ decays in VDM
\end{center}
%%%%%%%%%%%%%%%%%%%%%%%%%%%%%%%%%%%%%%%%%%%%%
We use the following interactions for $SVV$, $S'VV$ and $GVV$ coupling with 
coupling constants $B$, $B'$ and $B''$, respectively, 
\begin{equation}
L_I=B\varepsilon^{abc}\varepsilon_{def}S^d_a{V^{\mu\nu}}^e_b{V_{\mu\nu}}^f_c+
B'S'^b_a\{{V^{\mu\nu}}^c_b,\ {V_{\mu\nu}}^a_c\}+
B''G\{{V^{\mu\nu}}^b_a,\ {V_{\mu\nu}}^a_b\},
\end{equation}
where $\dis{V^{\mu\nu}}$ is the vector field strength $\dis{\partial^{\mu}V^\nu-
\partial^{\nu}V^\mu}$. 
These interactions are represented graphically by the diagrams shown in fig.~6. 
Although interactions as ${\rm Tr}(S{V_{\mu\nu})}{\rm Tr}(V^{\mu\nu})$ 
and ${\rm Tr}(S'V_{\mu\nu}){\rm Tr}(V^{\mu\nu})$ other than those 
represented by Eq.~(18) may exist \cite{schechter}, these 
interactions violate the $OZI$ rule and are considered to be smaller than the 
interactions in Eq.~(18). 
\vspace{0.0cm}\\
%%%%%%%%%%%%%%%%Fig.6%%%%%%%%%%%%%%%%%%%%%
\begin{center}
%WinTpicVersion2.15
\unitlength 0.1in
\begin{picture}(17.20,11.07)(2.60,-12.79)
% LINE 1 0 3 0
% 10 394 888 1184 888 1184 888 1975 572 394 1363 1184 1363 1184 1363 1975 1679 1975 1679 1975 1679
% 
\special{pn 13}%
\special{pa 394 488}%
\special{pa 1184 488}%
\special{fp}%
\special{pa 1184 488}%
\special{pa 1975 172}%
\special{fp}%
\special{pa 394 963}%
\special{pa 1184 963}%
\special{fp}%
\special{pa 1184 963}%
\special{pa 1975 1279}%
\special{fp}%
\special{pa 1975 1279}%
\special{pa 1975 1279}%
\special{fp}%
% LINE 1 0 3 0
% 10 394 1205 1026 1205 1026 1205 1975 825 394 1047 1026 1047 1026 1047 1184 1110 1248 1141 1975 1442
% 
\special{pn 13}%
\special{pa 394 805}%
\special{pa 1026 805}%
\special{fp}%
\special{pa 1026 805}%
\special{pa 1975 425}%
\special{fp}%
\special{pa 394 647}%
\special{pa 1026 647}%
\special{fp}%
\special{pa 1026 647}%
\special{pa 1184 710}%
\special{fp}%
\special{pa 1248 741}%
\special{pa 1975 1042}%
\special{fp}%
% VECTOR 1 0 3 0
% 2 394 888 710 888
% 
\special{pn 13}%
\special{pa 394 488}%
\special{pa 710 488}%
\special{fp}%
\special{sh 1}%
\special{pa 710 488}%
\special{pa 643 468}%
\special{pa 657 488}%
\special{pa 643 508}%
\special{pa 710 488}%
\special{fp}%
% VECTOR 1 0 3 0
% 2 386 1047 702 1047
% 
\special{pn 13}%
\special{pa 386 647}%
\special{pa 702 647}%
\special{fp}%
\special{sh 1}%
\special{pa 702 647}%
\special{pa 635 627}%
\special{pa 649 647}%
\special{pa 635 667}%
\special{pa 702 647}%
\special{fp}%
% VECTOR 1 0 3 0
% 2 1026 1205 647 1205
% 
\special{pn 13}%
\special{pa 1026 805}%
\special{pa 647 805}%
\special{fp}%
\special{sh 1}%
\special{pa 647 805}%
\special{pa 714 825}%
\special{pa 700 805}%
\special{pa 714 785}%
\special{pa 647 805}%
\special{fp}%
% VECTOR 1 0 3 0
% 2 963 1363 647 1363
% 
\special{pn 13}%
\special{pa 963 963}%
\special{pa 647 963}%
\special{fp}%
\special{sh 1}%
\special{pa 647 963}%
\special{pa 714 983}%
\special{pa 700 963}%
\special{pa 714 943}%
\special{pa 647 963}%
\special{fp}%
% VECTOR 1 0 3 0
% 2 1184 888 1580 730
% 
\special{pn 13}%
\special{pa 1184 488}%
\special{pa 1580 330}%
\special{fp}%
\special{sh 1}%
\special{pa 1580 330}%
\special{pa 1511 336}%
\special{pa 1530 350}%
\special{pa 1525 373}%
\special{pa 1580 330}%
\special{fp}%
% VECTOR 1 0 3 0
% 2 1248 1141 1572 1276
% 
\special{pn 13}%
\special{pa 1248 741}%
\special{pa 1572 876}%
\special{fp}%
\special{sh 1}%
\special{pa 1572 876}%
\special{pa 1518 832}%
\special{pa 1523 855}%
\special{pa 1503 869}%
\special{pa 1572 876}%
\special{fp}%
% VECTOR 1 0 3 0
% 2 1896 857 1540 999
% 
\special{pn 13}%
\special{pa 1896 457}%
\special{pa 1540 599}%
\special{fp}%
\special{sh 1}%
\special{pa 1540 599}%
\special{pa 1609 593}%
\special{pa 1590 579}%
\special{pa 1595 556}%
\special{pa 1540 599}%
\special{fp}%
% VECTOR 1 0 3 0
% 2 1873 1640 1533 1497
% 
\special{pn 13}%
\special{pa 1873 1240}%
\special{pa 1533 1097}%
\special{fp}%
\special{sh 1}%
\special{pa 1533 1097}%
\special{pa 1587 1141}%
\special{pa 1582 1118}%
\special{pa 1602 1104}%
\special{pa 1533 1097}%
\special{fp}%
% STR 2 0 3 0
% 3 1980 710 1980 810 2 0
% $V$
\put(19.8000,-4.1000){\makebox(0,0)[lb]{$V$}}%
% STR 2 0 3 0
% 3 1980 1510 1980 1610 2 0
% $V$
\put(19.8000,-12.1000){\makebox(0,0)[lb]{$V$}}%
% STR 2 0 3 0
% 3 260 1100 260 1200 2 0
% $S$
\put(2.6000,-8.0000){\makebox(0,0)[lb]{$S$}}%
\end{picture}%
\hspace{1cm}%WinTpicVersion2.15
\unitlength 0.1in
\begin{picture}(18.55,9.41)(2.20,-11.61)
% LINE 1 0 3 0
% 8 364 962 1220 962 1220 962 2075 620 364 1219 1220 1219 1220 1219 2075 1561
% 
\special{pn 13}%
\special{pa 364 562}%
\special{pa 1220 562}%
\special{fp}%
\special{pa 1220 562}%
\special{pa 2075 220}%
\special{fp}%
\special{pa 364 819}%
\special{pa 1220 819}%
\special{fp}%
\special{pa 1220 819}%
\special{pa 2075 1161}%
\special{fp}%
% LINE 1 0 3 0
% 4 2075 1347 1391 1082 1391 1082 2075 825
% 
\special{pn 13}%
\special{pa 2075 947}%
\special{pa 1391 682}%
\special{fp}%
\special{pa 1391 682}%
\special{pa 2075 425}%
\special{fp}%
% VECTOR 1 0 3 0
% 2 364 962 792 962
% 
\special{pn 13}%
\special{pa 364 562}%
\special{pa 792 562}%
\special{fp}%
\special{sh 1}%
\special{pa 792 562}%
\special{pa 725 542}%
\special{pa 739 562}%
\special{pa 725 582}%
\special{pa 792 562}%
\special{fp}%
% VECTOR 1 0 3 0
% 2 963 1219 723 1219
% 
\special{pn 13}%
\special{pa 963 819}%
\special{pa 723 819}%
\special{fp}%
\special{sh 1}%
\special{pa 723 819}%
\special{pa 790 839}%
\special{pa 776 819}%
\special{pa 790 799}%
\special{pa 723 819}%
\special{fp}%
% VECTOR 1 0 3 0
% 2 1220 962 1733 757
% 
\special{pn 13}%
\special{pa 1220 562}%
\special{pa 1733 357}%
\special{fp}%
\special{sh 1}%
\special{pa 1733 357}%
\special{pa 1664 363}%
\special{pa 1683 377}%
\special{pa 1679 400}%
\special{pa 1733 357}%
\special{fp}%
% VECTOR 1 0 3 0
% 2 2075 834 1767 937
% 
\special{pn 13}%
\special{pa 2075 434}%
\special{pa 1767 537}%
\special{fp}%
\special{sh 1}%
\special{pa 1767 537}%
\special{pa 1837 535}%
\special{pa 1818 520}%
\special{pa 1824 497}%
\special{pa 1767 537}%
\special{fp}%
% VECTOR 1 0 3 0
% 2 2075 1561 1690 1407
% 
\special{pn 13}%
\special{pa 2075 1161}%
\special{pa 1690 1007}%
\special{fp}%
\special{sh 1}%
\special{pa 1690 1007}%
\special{pa 1744 1050}%
\special{pa 1740 1027}%
\special{pa 1759 1013}%
\special{pa 1690 1007}%
\special{fp}%
% VECTOR 1 0 3 0
% 2 1519 1125 1810 1245
% 
\special{pn 13}%
\special{pa 1519 725}%
\special{pa 1810 845}%
\special{fp}%
\special{sh 1}%
\special{pa 1810 845}%
\special{pa 1756 801}%
\special{pa 1761 825}%
\special{pa 1741 838}%
\special{pa 1810 845}%
\special{fp}%
% STR 2 0 3 0
% 3 2070 710 2070 810 2 0
% $V$
\put(20.7000,-4.1000){\makebox(0,0)[lb]{$V$}}%
% STR 2 0 3 0
% 3 2070 1410 2070 1510 2 0
% $V$
\put(20.7000,-11.1000){\makebox(0,0)[lb]{$V$}}%
% STR 2 0 3 0
% 3 220 1090 220 1190 2 0
% $S'$
\put(2.2000,-7.9000){\makebox(0,0)[lb]{$S'$}}%
\end{picture}%
\hspace{1cm}%WinTpicVersion3.08
\unitlength 0.1in
\begin{picture}( 17.7000, 10.6000)(  2.6000,-12.5000)
% FUNC 1 0 3 0
% 9 400 400 1200 800 400 600 600 600 400 400 400 400 1200 800 0 3 3 0
% 1/2*sin(4*x)
\special{pn 13}%
\special{pa 400 600}%
\special{pa 406 590}%
\special{pa 410 582}%
\special{pa 416 574}%
\special{pa 420 570}%
\special{pa 426 568}%
\special{pa 430 570}%
\special{pa 436 574}%
\special{pa 440 582}%
\special{pa 446 590}%
\special{pa 450 600}%
\special{pa 456 610}%
\special{pa 460 620}%
\special{pa 466 626}%
\special{pa 470 630}%
\special{pa 476 632}%
\special{pa 480 630}%
\special{pa 486 626}%
\special{pa 490 620}%
\special{pa 496 610}%
\special{pa 500 600}%
\special{pa 506 590}%
\special{pa 510 582}%
\special{pa 516 574}%
\special{pa 520 570}%
\special{pa 526 568}%
\special{pa 530 570}%
\special{pa 536 574}%
\special{pa 540 582}%
\special{pa 546 590}%
\special{pa 550 600}%
\special{pa 556 610}%
\special{pa 560 620}%
\special{pa 566 626}%
\special{pa 570 630}%
\special{pa 576 632}%
\special{pa 580 630}%
\special{pa 586 626}%
\special{pa 590 620}%
\special{pa 596 610}%
\special{pa 600 600}%
\special{pa 606 590}%
\special{pa 610 582}%
\special{pa 616 574}%
\special{pa 620 570}%
\special{pa 626 568}%
\special{pa 630 570}%
\special{pa 636 574}%
\special{pa 640 582}%
\special{pa 646 590}%
\special{pa 650 600}%
\special{pa 656 610}%
\special{pa 660 620}%
\special{pa 666 626}%
\special{pa 670 630}%
\special{pa 676 632}%
\special{pa 680 630}%
\special{pa 686 626}%
\special{pa 690 620}%
\special{pa 696 610}%
\special{pa 700 600}%
\special{pa 706 590}%
\special{pa 710 582}%
\special{pa 716 574}%
\special{pa 720 570}%
\special{pa 726 568}%
\special{pa 730 570}%
\special{pa 736 574}%
\special{pa 740 582}%
\special{pa 746 590}%
\special{pa 750 600}%
\special{pa 756 610}%
\special{pa 760 620}%
\special{pa 766 626}%
\special{pa 770 630}%
\special{pa 776 632}%
\special{pa 780 630}%
\special{pa 786 626}%
\special{pa 790 620}%
\special{pa 796 610}%
\special{pa 800 600}%
\special{pa 806 590}%
\special{pa 810 582}%
\special{pa 816 574}%
\special{pa 820 570}%
\special{pa 826 568}%
\special{pa 830 570}%
\special{pa 836 574}%
\special{pa 840 582}%
\special{pa 846 590}%
\special{pa 850 600}%
\special{pa 856 610}%
\special{pa 860 620}%
\special{pa 866 626}%
\special{pa 870 630}%
\special{pa 876 632}%
\special{pa 880 630}%
\special{pa 886 626}%
\special{pa 890 620}%
\special{pa 896 610}%
\special{pa 900 600}%
\special{pa 906 590}%
\special{pa 910 582}%
\special{pa 916 574}%
\special{pa 920 570}%
\special{pa 926 568}%
\special{pa 930 570}%
\special{pa 936 574}%
\special{pa 940 582}%
\special{pa 946 590}%
\special{pa 950 600}%
\special{pa 956 610}%
\special{pa 960 620}%
\special{pa 966 626}%
\special{pa 970 630}%
\special{pa 976 632}%
\special{pa 980 630}%
\special{pa 986 626}%
\special{pa 990 620}%
\special{pa 996 610}%
\special{pa 1000 600}%
\special{pa 1006 590}%
\special{pa 1010 582}%
\special{pa 1016 574}%
\special{pa 1020 570}%
\special{pa 1026 568}%
\special{pa 1030 570}%
\special{pa 1036 574}%
\special{pa 1040 582}%
\special{pa 1046 590}%
\special{pa 1050 600}%
\special{pa 1056 610}%
\special{pa 1060 620}%
\special{pa 1066 626}%
\special{pa 1070 630}%
\special{pa 1076 632}%
\special{pa 1080 630}%
\special{pa 1086 626}%
\special{pa 1090 620}%
\special{pa 1096 610}%
\special{pa 1100 600}%
\special{pa 1106 590}%
\special{pa 1110 582}%
\special{pa 1116 574}%
\special{pa 1120 570}%
\special{pa 1126 568}%
\special{pa 1130 570}%
\special{pa 1136 574}%
\special{pa 1140 582}%
\special{pa 1146 590}%
\special{pa 1150 600}%
\special{pa 1156 610}%
\special{pa 1160 620}%
\special{pa 1166 626}%
\special{pa 1170 630}%
\special{pa 1176 632}%
\special{pa 1180 630}%
\special{pa 1186 626}%
\special{pa 1190 620}%
\special{pa 1196 610}%
\special{pa 1200 600}%
\special{sp}%
% FUNC 1 0 3 0
% 9 400 650 1200 1050 400 850 600 850 400 650 400 650 1200 1050 0 3 3 0
% 1/2*sin(4*x)
\special{pn 13}%
\special{pa 400 850}%
\special{pa 406 840}%
\special{pa 410 832}%
\special{pa 416 824}%
\special{pa 420 820}%
\special{pa 426 818}%
\special{pa 430 820}%
\special{pa 436 824}%
\special{pa 440 832}%
\special{pa 446 840}%
\special{pa 450 850}%
\special{pa 456 860}%
\special{pa 460 870}%
\special{pa 466 876}%
\special{pa 470 880}%
\special{pa 476 882}%
\special{pa 480 880}%
\special{pa 486 876}%
\special{pa 490 870}%
\special{pa 496 860}%
\special{pa 500 850}%
\special{pa 506 840}%
\special{pa 510 832}%
\special{pa 516 824}%
\special{pa 520 820}%
\special{pa 526 818}%
\special{pa 530 820}%
\special{pa 536 824}%
\special{pa 540 832}%
\special{pa 546 840}%
\special{pa 550 850}%
\special{pa 556 860}%
\special{pa 560 870}%
\special{pa 566 876}%
\special{pa 570 880}%
\special{pa 576 882}%
\special{pa 580 880}%
\special{pa 586 876}%
\special{pa 590 870}%
\special{pa 596 860}%
\special{pa 600 850}%
\special{pa 606 840}%
\special{pa 610 832}%
\special{pa 616 824}%
\special{pa 620 820}%
\special{pa 626 818}%
\special{pa 630 820}%
\special{pa 636 824}%
\special{pa 640 832}%
\special{pa 646 840}%
\special{pa 650 850}%
\special{pa 656 860}%
\special{pa 660 870}%
\special{pa 666 876}%
\special{pa 670 880}%
\special{pa 676 882}%
\special{pa 680 880}%
\special{pa 686 876}%
\special{pa 690 870}%
\special{pa 696 860}%
\special{pa 700 850}%
\special{pa 706 840}%
\special{pa 710 832}%
\special{pa 716 824}%
\special{pa 720 820}%
\special{pa 726 818}%
\special{pa 730 820}%
\special{pa 736 824}%
\special{pa 740 832}%
\special{pa 746 840}%
\special{pa 750 850}%
\special{pa 756 860}%
\special{pa 760 870}%
\special{pa 766 876}%
\special{pa 770 880}%
\special{pa 776 882}%
\special{pa 780 880}%
\special{pa 786 876}%
\special{pa 790 870}%
\special{pa 796 860}%
\special{pa 800 850}%
\special{pa 806 840}%
\special{pa 810 832}%
\special{pa 816 824}%
\special{pa 820 820}%
\special{pa 826 818}%
\special{pa 830 820}%
\special{pa 836 824}%
\special{pa 840 832}%
\special{pa 846 840}%
\special{pa 850 850}%
\special{pa 856 860}%
\special{pa 860 870}%
\special{pa 866 876}%
\special{pa 870 880}%
\special{pa 876 882}%
\special{pa 880 880}%
\special{pa 886 876}%
\special{pa 890 870}%
\special{pa 896 860}%
\special{pa 900 850}%
\special{pa 906 840}%
\special{pa 910 832}%
\special{pa 916 824}%
\special{pa 920 820}%
\special{pa 926 818}%
\special{pa 930 820}%
\special{pa 936 824}%
\special{pa 940 832}%
\special{pa 946 840}%
\special{pa 950 850}%
\special{pa 956 860}%
\special{pa 960 870}%
\special{pa 966 876}%
\special{pa 970 880}%
\special{pa 976 882}%
\special{pa 980 880}%
\special{pa 986 876}%
\special{pa 990 870}%
\special{pa 996 860}%
\special{pa 1000 850}%
\special{pa 1006 840}%
\special{pa 1010 832}%
\special{pa 1016 824}%
\special{pa 1020 820}%
\special{pa 1026 818}%
\special{pa 1030 820}%
\special{pa 1036 824}%
\special{pa 1040 832}%
\special{pa 1046 840}%
\special{pa 1050 850}%
\special{pa 1056 860}%
\special{pa 1060 870}%
\special{pa 1066 876}%
\special{pa 1070 880}%
\special{pa 1076 882}%
\special{pa 1080 880}%
\special{pa 1086 876}%
\special{pa 1090 870}%
\special{pa 1096 860}%
\special{pa 1100 850}%
\special{pa 1106 840}%
\special{pa 1110 832}%
\special{pa 1116 824}%
\special{pa 1120 820}%
\special{pa 1126 818}%
\special{pa 1130 820}%
\special{pa 1136 824}%
\special{pa 1140 832}%
\special{pa 1146 840}%
\special{pa 1150 850}%
\special{pa 1156 860}%
\special{pa 1160 870}%
\special{pa 1166 876}%
\special{pa 1170 880}%
\special{pa 1176 882}%
\special{pa 1180 880}%
\special{pa 1186 876}%
\special{pa 1190 870}%
\special{pa 1196 860}%
\special{pa 1200 850}%
\special{sp}%
% VECTOR 1 0 3 0
% 2 1200 600 1600 400
% 
\special{pn 13}%
\special{pa 1200 600}%
\special{pa 1600 400}%
\special{fp}%
\special{sh 1}%
\special{pa 1600 400}%
\special{pa 1532 412}%
\special{pa 1552 424}%
\special{pa 1550 448}%
\special{pa 1600 400}%
\special{fp}%
% LINE 1 0 3 0
% 2 1600 400 2000 200
% 
\special{pn 13}%
\special{pa 1600 400}%
\special{pa 2000 200}%
\special{fp}%
% VECTOR 1 0 3 0
% 2 2000 1250 1600 1050
% 
\special{pn 13}%
\special{pa 2000 1250}%
\special{pa 1600 1050}%
\special{fp}%
\special{sh 1}%
\special{pa 1600 1050}%
\special{pa 1652 1098}%
\special{pa 1648 1074}%
\special{pa 1670 1062}%
\special{pa 1600 1050}%
\special{fp}%
% LINE 1 0 3 0
% 2 1590 1050 1190 850
% 
\special{pn 13}%
\special{pa 1590 1050}%
\special{pa 1190 850}%
\special{fp}%
% LINE 1 0 3 0
% 2 1200 600 1200 850
% 
\special{pn 13}%
\special{pa 1200 600}%
\special{pa 1200 850}%
\special{fp}%
% LINE 1 0 3 0
% 2 1400 720 1700 570
% 
\special{pn 13}%
\special{pa 1400 720}%
\special{pa 1700 570}%
\special{fp}%
% VECTOR 1 0 3 0
% 2 2000 410 1700 570
% 
\special{pn 13}%
\special{pa 2000 410}%
\special{pa 1700 570}%
\special{fp}%
\special{sh 1}%
\special{pa 1700 570}%
\special{pa 1768 556}%
\special{pa 1748 546}%
\special{pa 1750 522}%
\special{pa 1700 570}%
\special{fp}%
% VECTOR 1 0 3 0
% 2 1400 720 1760 900
% 
\special{pn 13}%
\special{pa 1400 720}%
\special{pa 1760 900}%
\special{fp}%
\special{sh 1}%
\special{pa 1760 900}%
\special{pa 1710 852}%
\special{pa 1712 876}%
\special{pa 1692 888}%
\special{pa 1760 900}%
\special{fp}%
% LINE 1 0 3 0
% 2 1750 900 2000 1030
% 
\special{pn 13}%
\special{pa 1750 900}%
\special{pa 2000 1030}%
\special{fp}%
% STR 2 0 3 0
% 3 260 690 260 790 2 0
% $G$
\put(2.6000,-7.9000){\makebox(0,0)[lb]{$G$}}%
% STR 2 0 3 0
% 3 2030 260 2030 360 2 0
% $V$
\put(20.3000,-3.6000){\makebox(0,0)[lb]{$V$}}%
% STR 2 0 3 0
% 3 2030 1080 2030 1180 2 0
% $V$
\put(20.3000,-11.8000){\makebox(0,0)[lb]{$V$}}%
\end{picture}%
\\
Fig.6. $SVV$, $S'VV$ and $GVV$ coupling
\end{center}
%%%%%%%%%%%%%%%%%%%%%%%%%%%%%%%%%%%%%%%%%%%%%
\par
We define the coupling constants $g_{SVV'}$ in the 
following expression, 
\begin{eqnarray}
L_I&=&g_{a_0K^*\overline{K^*}}\frac{1}{\sqrt{2}}\overline{K^*_{\mu\nu}}
\bi{\tau}\bi{\cdot}\bi{a_0}{K^*}^{\mu\nu}
+g_{a'_0K^*\overline{K^*}}\frac{1}{\sqrt{2}}\overline{K^*_{\mu\nu}}
\bi{\tau}\bi{\cdot}\bi{a'_0}{K^*}^{\mu\nu}
+g_{a_0\rho\omega}\bi{a_0\cdot}\bi{\rho}_{\mu\nu}\omega^{\mu\nu}\nonumber\\
&+&g_{a'_0\rho\omega}\bi{a'_0\cdot}\bi{\rho}_{\mu\nu}\omega^{\mu\nu}
+g_{a_0\rho\phi}\bi{a_0\cdot}\bi{\rho}_{\mu\nu}\phi^{\mu\nu}
+g_{a'_0\rho\phi}\bi{a'_0\cdot}\bi{\rho}_{\mu\nu}\phi^{\mu\nu}\nonumber\\
&+&g_{K^*_0 K^*\rho}(\frac{1}{\sqrt{2}}\overline{K^*_{\mu\nu}}\bi{\tau\cdot}
\bi{\rho}^{\mu\nu}K_0^*+H.C.)
+g_{\kappa K^*\rho}(\frac{1}{\sqrt{2}}\overline{K^*_{\mu\nu}}\bi{\tau\cdot}
\bi{\rho}^{\mu\nu}\kappa+H.C.)\nonumber\\
&+&g_{\kappa K^*\omega}(\overline{\kappa}K^*_{\mu\nu}\omega^{\mu\nu}+H.C.)
+g_{K^*_0 K^*\omega}(\overline{K^*_0}K^*_{\mu\nu}\omega^{\mu\nu}+H.C.)\nonumber\\
&+&g_{\kappa K^*\phi}(\overline{\kappa}K^*_{\mu\nu}\phi^{\mu\nu}
+H.C.)
+g_{K^*_0K^*\phi}(\overline{K^*_0}K^*_{\mu\nu}\phi^{\mu\nu}
+H.C.),\nonumber\\
&+&g_{f_0(M)\rho\rho}f_0(M)\bi{\rho}_{\mu\nu}{\cdot}
\bi{\rho}^{\mu\nu}
+g_{f_0(M)K^*\overline{K^*}}f_0(M)K^*_{\mu\nu}\overline{K^*}^{\mu\nu}
+g_{f_0(M)\omega\omega}f_0(M)\omega_{\mu\nu}\omega^{\mu\nu}
\nonumber\\
&+&g_{f_0(M)\omega\phi}f_0(M)\omega_{\mu\nu}\phi^{\mu\nu}
+g_{f_0(M)\phi\phi}f_0(M)\phi_{\mu\nu}\phi^{\mu\nu},
\end{eqnarray}
where fields $a_0$ represents the low mass $I=1$ scalar mesons and $a_0'$ 
the high mass $I=1$ scalar mesons. Then the  coupling constants for $I_3=0$ $S$, 
$V$ mesons are, by using Eq.~(18), expressed  as
\begin{eqnarray}
g_{a_0(980)\rho\phi}&=&-2B\cos\theta_a,\nonumber\\
g_{a_0(980)\rho\omega}&=&-2\sqrt{2}B'\sin\theta_a ,\nonumber\\
g_{f_0(980)\phi\omega}&=&2BR_{f_0(980)NS},\nonumber\\
g_{f_0(980)\phi\phi}&=&2B'R_{f_0(980)S'}+2B''R_{f_0(980)G},\nonumber\\
g_{f_0(980)\rho\rho}&=&-BR_{f_0(980)NN}+\sqrt{2}B'R_{f_0(980)N'}
    +2B''R_{f_0(980)G},\nonumber\\
g_{f_0(980)\omega\omega}&=&BR_{f_0(980)NN}+\sqrt{2}B'R_{f_0(980)N'}
    +2B''R_{f_0(980)G}.
\end{eqnarray}
$V-\gamma$ coupling is defined as 
\begin{equation}
<0|j_{\mu}^{\rm em}(0)|V(p,\varepsilon)>=\frac{em_V^2}{\gamma_{V}}
\varepsilon_\mu(p),
\end{equation}
where $m_V$ and $\varepsilon_\mu(p)$ are the mass and polarization 
vector of the vector meson, respectively. Then $\dis{\Gamma(V\to ll)=
\frac{4\pi}{3}\frac{\alpha^2}{\gamma_V^2}m_V}$, we can get the value 
$\dis{\frac1\gamma_{\rho}=0.201\pm0.002}$ from the experimental data 
$\Gamma(\rho\to ee)=7.02\pm0.11{\rm keV}$ \cite{PDG}.
\par
The decay amplitude and decay width for $V\to S\gamma$ are expressed as 
\begin{eqnarray}
M(V\to S\gamma)&=&\sum_{V'}\frac{2eg_{SVV'}}{\gamma_{V'}}(p\cdot k\varepsilon_V\cdot
\varepsilon_{\gamma}-p\cdot\varepsilon_{\gamma}k\cdot\varepsilon_V),\nonumber\\
\Gamma(V\to S\gamma)&=&\frac{4\alpha}{3}(\sum_{V'}\frac{g_{SVV'}}{\gamma_{V'}})^2
|{\bi k}_{\gamma}|^3,
\end{eqnarray}
where $\varepsilon_V$ and $\varepsilon_{\gamma}$ are the polarization vector of 
the vector meson and the photon, respectively and ${\bi k}_{\gamma}$ is 
the photon momentum in the $V$ rest frame. Only the $\rho$ meson contributes to 
the intermediate $V'$ vector meson for the $\phi\to a_0(980)\gamma$ decay, and 
$\omega$ and $\phi$ mesons contribute to $\phi\to f_0(980)\gamma$ decay. Then the 
decay widths for the $\phi\to a_0(980)\gamma$ and $\phi\to f_0(980)\gamma$ decay 
are written as
\begin{eqnarray}
\Gamma(\phi\to a_0(980)\gamma)&=&\frac{4\alpha}{3\gamma_{\rho}^2}
(g_{a_0\rho\phi})^2|{\bi k}_{\gamma}|^3,\nonumber\\
\Gamma(\phi\to f_0(980)\gamma)&=&\frac{4\alpha}{3\gamma_{\rho}^2}
(\frac{g_{f_0\phi\omega}}{3}-\frac{\sqrt{2}g_{f_0\phi\phi}}{3})^2
|{\bi k}_{\gamma}|^3,
\end{eqnarray}
where we assumed the $SU(3)$ symmetry for the $V-\gamma$ coupling,  
$$
\frac{m_{\rho}^2}{\gamma_{\rho}}:\frac{m_{\omega}^2}{\gamma_{\omega}}:
\frac{m_{\phi}^2}{\gamma_{\phi}}\cong\frac{1}{\gamma_{\rho}}:\frac{1}
{\gamma_{\omega}}:\frac{1}{\gamma_{\phi}}=\frac{1}{\sqrt{2}}:\frac{1}{3\sqrt{2}}:
-\frac{1}{3}.
$$
For the decay $S\to\gamma\gamma$, decay amplitude and width are expressed as 
\begin{eqnarray}
M(S\to \gamma\gamma)&=&\sum_{VV'}\frac{2e^2g_{SVV'}}{\gamma_{V}\gamma_{V'}}
(k_1\cdot k_2\varepsilon_1\cdot\varepsilon_2-k_1\cdot\varepsilon_2k_2\cdot
\varepsilon_1),\nonumber\\
\Gamma(S\to \gamma\gamma)&=&{\pi\alpha^2}m_S^3(\sum_{VV'}\frac{g_{SVV'}}
{\gamma_{V}\gamma_{V'}})^2,
\end{eqnarray}
where $k_1$, $k_2$ and $\varepsilon_1$, $\varepsilon_2$ are the photon momentums 
and polarization vectors of photon. 
The decay widths for $a_0(980)\to\gamma\gamma$ and $f_0(980)\to\gamma\gamma$ 
are expressed as 
\begin{eqnarray}
\Gamma(a_0(980)\to\gamma\gamma)&=&\frac{\pi\alpha^2}{\gamma_{\rho}^2}m_{a_0}^3
(-\frac{\sqrt{2}}{3}g_{a_0\rho\phi}+\frac13g_{a_0\rho\omega})^2,\nonumber\\
\Gamma(f_0(980)\to\gamma\gamma)&=&\frac{\pi\alpha^2}{\gamma_{\rho}^2}m_{f_0}^3
(-\frac{\sqrt{2}}{9}g_{f_0\phi\omega}+g_{f_0\rho\rho}+\frac{1}{9}
g_{f_0\omega\omega}+\frac{2}{9}g_{f_0\phi\phi})^2.
\end{eqnarray}
\subsection{Numerical Analysis}
We quote the experimental data for these decay widths from PDG \cite{PDG},
\begin{eqnarray}
\Gamma(\phi\to a_0(980)\gamma)&=&0.323\pm0.029\ {\rm keV},\nonumber\\
\Gamma(\phi\to f_0(980)\gamma)&=&1.87\pm0.11\ {\rm keV},\nonumber\\
\Gamma(a_0(980)\to\gamma\gamma)&=&0.30\pm0.10\ {\rm keV},\nonumber\\
\Gamma(f_0(980)\to\gamma\gamma)&=&0.39\makebox{$ {+0.10\atop-0.13}$}\ {\rm keV}.
\end{eqnarray} 
The coupling strengths for these radiative decays are described by the coupling 
constants $B$, $B'$ abd $B''$ from the equations (20), (23) and (25), and 
the values for these coupling strengths are determined by the experimental data 
Eq. (26) as
\begin{eqnarray}
|B\cos\theta_a|&=&2.316\pm0.127\ {\rm GeV}^{-1},\ {\rm from}\ 
\phi\to a_0(980)\gamma\ {\rm decay}\nonumber\\
|BR_{f_0NS}-\sqrt{2}B'R_{f_0S'}-\sqrt{2}B''R_{f_0G}|&=&13.8\pm0.54\ {\rm GeV}^{-1},
\ {\rm from}\ \phi\to f_0(980)\gamma\ {\rm decay}\nonumber\\
|B\cos\theta_a-B'\sin{\theta_a}|&=&1.137\pm0.214\ {\rm GeV}^{-1},
\ {\rm from}\ a_0(980)\to\gamma\gamma\ {\rm decay}\nonumber\\
|B(R_{f_0NS}+2\sqrt{2}R_{f_0NN})-B'(5R_{f_0N'}&+&\sqrt{2}R_{f_0S'})-6\sqrt{2}
B''R_{f_0G}|=
3.920\makebox{$ {+0.591\atop-0.789}$}\ {\rm GeV}^{-1},\nonumber\\
&& {\rm from}\ f_0(980)\to\gamma\gamma\ {\rm decay.}
\end{eqnarray}
\par 
Because only the 4 quark component of $a_0(980)$ can contribute to the $\phi\to 
a_0(980)\gamma$ decay,  one can not explain the $\phi\to a_0(980)\gamma$ decay
width without the 4-quark model for the low mass scalar mesons.
The experimental fact that the coupling strength of $\phi\to f_0(980)\gamma$ 
decay is larger as several factor than that of $\phi\to a_0(980)\gamma$ can be  
explained by the mixing of the 2 quark state in the $f_0(980)$ meson, $\sqrt{2}
B'R_{f_0S'}$ factor. If we take the relative sign of the $B'$ to the $B$ as 
same, $\sqrt{2}B'R_{f_0S'}$ factor contributes additively by the opposite sign 
of the $R_{f_0NS}$ to the $R_{f_0S'}$ (see Eq. (17)). The coupling strength for 
$a_0(980)\to \gamma\gamma$ is a half of that for $\phi\to f_0(980)\gamma$. This 
can be explained easily by the fact that $B$ and $B'$ have the same sign and 
$\theta_a$ is positive, then $B'\sin\theta_a$ term contribute destructively.
The term $BR_{f_0NS}-\sqrt{2}B'R_{f_0S'}-\sqrt{2}B''R_{f_0G}$ in the coupling of $
f_0(980)\to\gamma\gamma$ decay is the same as that in $\phi\to f_0(980)\gamma$ 
decay. The term $2\sqrt{2}BR_{f_0NN}$ in the coupling of $f_0(980)\to\gamma
\gamma$ decay contributes destructively because of the opposite sign of the mixing  
parameter $R_{f_0NN}$ to that of the $R_{f_0NS}$ (see the eq. (17)). 
\par
We got the best fit values of coupling constant $B$, $B'$ and $B''$ taking the 
$\chi^2$ fit analysis for the experimental data and these coupling constants 
contained in Eq. (27).  
Because the coupling strengths contain the mixing parameters $R_{a_0NS}$ etc., 
we vary the values of $R_{f_0NS}$ etc. in the $\chi^2$ analysis. 
Variations of the mixing parameters $R_{a_0NS}$ etc. are caused by the shift of 
the values of $\lambda_{01}$, $\lambda_{0}$, $\lambda_{1}$, $\lambda_{GG}$, 
$\lambda_{G}$, then we perform the $\chi^2$ fit varying these parameters. 
The best fit values of the coupling strengths for radiative 
decays are obtained and shown in Table I. These values are given on the following 
values of coupling constants $B$ etc. and mixing mass parameters $\lambda_{01}$ 
etc.; 
\begin{eqnarray}
&&B=2.8{\rm GeV^{-1}},\ B'=12.0{\rm GeV^{-1}},\ B''=7.2{\rm GeV^{-1}}, \nonumber\\
&&\lambda_{01}=0.24{\rm GeV^2},\ \lambda_{0}=-0.03{\rm GeV^2},\ 
\lambda_{1}=0.09{\rm GeV^2},\nonumber\\
&&\hspace{2cm} \lambda_{G}=0.17{\rm GeV^2},\ \lambda_{GG}=
1.66^2{\rm GeV^2}.
\end{eqnarray}
%%%%%%%%%%%%%%%%%%%%Table I%%%%%%%%%%%%%%%%%%%%%%%%%%%%
\begin{table}
\caption{Experimental data \cite{PDG} and best-fit values of coupling strength 
for radiative decay involving $a_0(980)$ and $f_0(980)$. 
These best-fit values are obtained on the values of coupling and mass parameters; 
$B=2.8{\rm GeV^{-1}}$, $B'=12.0{\rm GeV^{-1}}$, $B''=7.2{\rm GeV^{-1}}$, 
$\lambda_{01}=0.24{\rm GeV^2}$, $\lambda_{0}=-0.03{\rm GeV^2}$, $\lambda_{1}=
0.09{\rm GeV^2}$, $\lambda_{G}=0.17{\rm GeV^2}$, $\lambda_{GG}=1.66^2{\rm GeV^2}$.
}  
\begin{center}
\begin{tabular}{|c|c|c|c|}\hline\hline
Decay&Coupling strength &Experimental data &Best-fit value\\ \hline
$\phi\to a_0(980)\gamma$&$|B\cos\theta_a|$ &$2.316\pm0.127\ {\rm GeV}^{-1}$&
$2.76\ {\rm GeV}^{-1}$\\
$\phi\to f_0(980)\gamma$&$|BR_{f_0NS}-\sqrt{2}B'R_{f_0S'}-\sqrt{2}B''R_{f_0G}|$&$
13.8\pm0.54\ {\rm GeV}^{-1}$&$7.15\ {\rm GeV}^{-1}$\\
$a_0(980)\to \gamma\gamma$&$|B\cos\theta_a-B'\sin{\theta_a}|$&$1.137\pm0.214\ 
{\rm GeV}^{-1}$&$0.84\ {\rm GeV}^{-1}$\\
$f_0(980)\to\gamma\gamma$&$|B(R_{f_0NS}+2\sqrt{2}R_{f_0NN})-B'(5R_{f_0N'}$&&\\
&$+\sqrt{2}R_{f_0S'})-6\sqrt{2}
B''R_{f_0G}|$&$3.920\makebox{$ {+0.591\atop-0.789}$}\ {\rm GeV}^{-1}$&$6.20\ 
{\rm GeV}^{-1}$ \\ \hline\hline
\end{tabular}
\end{center}
\end{table}
%%%%%%%%%%%%%%%%%%%%%%%%%%%%%%%%%%%%%%%%%%%%%%%
At this time, obtained mass eigenvalues and mixing parameters for $f_0(980)$ etc. 
are as follows;
\begin{eqnarray}
&&m_{f_0(600)}=0.748{\rm GeV}, \ m_{f_0(980)}=0.912{\rm GeV}, \  m_{f_0(1370)}=
1.381{\rm GeV},\nonumber\\
&&m_{f_0(1500)}=1.522{\rm GeV},\ m_{f_0(1710)}=1.711{\rm GeV},\nonumber\\
&&\left(\begin{array}{c}
    f_0(600)\\
    f_0(980)\\
    f_0(1370)\\
    f_0(1500)\\
    f_0(1710)
    \end{array}\right)=
    \left(\begin{array}{ccccc}
     0.929&0.292&-0.218&-0.058&0.028\\
     0.315&-0.901&0.053&0.292&-0.032\\
     0.088&-0.251&0.295&-0.912& 0.099\\
     0.160&0.178&0.808&0.172&-0.510\\
     0.067&0.092&0.458&0.222&0.853     
     \end{array}\right)
    \left(\begin{array}{c}
    f_{NN}\\
    f_{NS}\\
    f_{N'}\\
    f_{S'}\\
    f_G
    \end{array}\right).
\end{eqnarray}
\par
We can explain the radiative decays, $\phi\to a_0(980)\gamma$, $\phi\to f_0(980)
\gamma$, $a_0(980)\to\gamma\gamma$ and $f_0(980)\to\gamma\gamma$ comprehensively 
in the VDM model, assuming that the low mass scalar $a_0(980)$ and 
$f_0(980)$ mesons are composed of $qq\bar{q}\bar{q}$ state dominantly and there 
exists the mixing between the low mass scalar and high mass $q\bar{q}$ scalar 
mesons. 
The literature \cite{phi decay II} studied these decays in the VDM systematically, 
but it could not explain the experimental large decay width of $\phi\to 
f_0(980)\gamma$. 
To explain this large decay width 
of $\phi\to f_0(980)\gamma$, the work \cite{CLOSE} applied the $a_0$-$f_0$ mixing. 
But that the $a_0$-$f_0$ mixing effect for the $\phi\to f_0(980)\gamma$ decay is 
small was shown by the authors N. N. Achasov and A. V. Kiselev \cite{ACHASOV}. 
We explain this 
large decay width of $\phi\to f_0(980)\gamma$ by the $qq\bar{q}\bar{q}$ 
scalar and $q\bar{q}$ scalar meson mixing. 
\par
Although we can make the coupling strength for $\phi\to f_0(980)\gamma$ so large, 
the best-fit value of this is a half of the experimental data. On the other hand, 
the best fit value of the coupling strength for $f_0(980)\to\gamma\gamma$ is 1.6 
times of the experimental data. In our model, the coupling strengths 
$BR_{f_0NS}-\sqrt{2}B'R_{f_0S'}-\sqrt{2}B''R_{f_0G}$ are common in the couplings of 
$\phi\to f_0(980)\gamma$ and $f_0(980)\to\gamma\gamma$ decay and the residual term 
$2\sqrt{2}BR_{f_0NN}-5B'R_{f_0N'}-5\sqrt{2}B''R_{f_0G}$ in the coupling for 
$f_0(980)\to\gamma\gamma$ are not so large, then the best fit values of these 
radiative decays are not so different. 
In order to explain the rather large difference of experimental data between 
$\phi\to f_0(980)\gamma$ and $f_0(980)\to\gamma\gamma$ decay, any other process 
than that deduced from the VDM may be considered. 
\par
In our present reanalysis for radiative decays of scalar mesons, we can get the 
best fit values for mixing parameters $R_{f_0NS}$ etc. and mass eigenvalues for 
$f_0(980)$ meson etc. Obtained mass values $0.912{\rm GeV}$ for $f_0(980)$ is 
rather small compared with the experimental value $0.980\pm0.010{\rm GeV}$.     
This mass value of $f_0(980)$ is affected with the input values of the masses for 
$m_{f_0(600)}$ and $m_{\kappa(900)}$ and these input masses are very uncertain at 
present, then we may get the mass eigenvalue for $f_0(980)$ nearer to the 
experimental values $0.980\pm0.010{\rm GeV}$ by sifting the input mass values 
for $f_0(600)$ and $\kappa(900)$.
%%%%%%%%%%%%%% section 4 %%%%%%%%%%%%%%%%%%%%%%%%
\section{Conclusion}
We analyzed the radiative decays $\phi\to f_0(980)\gamma$, $\phi\to a_0(980)
\gamma$, $f_0(980)\to\gamma\gamma$ and $a_0(980)\to\gamma\gamma$ using the VDM 
in the framework where $f_0(980)$ and $a_0(980)$ are mainly $qq\bar{q}\bar{q}$ 
scalar mesons and are mixed with $q\bar{q}$ high mass scalar mesons. 
Mixing between $a_0(980)$ and $a_0(1450)$ has been studied in our previous paper 
\cite{teshima2}, where decay processes of $a_0(980)$ and $a_0(1450)$ decaying to 
two pseudoscalar mesons are analyzed, and the mixing angle $\theta_a$ has 
estimated to be about $9^{\circ}$. 
The mass mixing parameter $\lambda_{12}$ causing the mixing 
between $qq\bar{q}\bar{q}$ state and $q\bar{q}$ state corresponds to values to be 
$0.19{\rm GeV^{2}}$. In present work, we estimated the mixing parameters 
$R_{f_0(980)NS}$ etc. performing the $\chi^2$ analysis varying the $\lambda_{12}$ 
near the values $\sim0.19{\rm GeV^{2}}$. 
\par
We assumed the form of the $SVV$ coupling $B\varepsilon^{abc}\varepsilon_{def}
S^d_a{V^{\mu\nu}}^e_b{V_{\mu\nu}}^f_c+B'S'^b_a\{{V^{\mu\nu}}^c_b,\ 
{V_{\mu\nu}}^a_c\}+B''G\{{V^{\mu\nu}}^b_a,\ {V_{\mu\nu}}^a_b\}$ similar to that 
for $SPP$ coupling, which was used in our previous work \cite{teshima2}. 
By assuming the VDM, the coupling strengths for $\phi\to a_0(980)\gamma$, $\phi\to 
f_0(980)\gamma$, $a_0(980)\to\gamma\gamma$ and $f_0(980)\to\gamma\gamma$ decays 
are expressed as 
$|B\cos\theta_a|$, 
$|BR_{f_0NS}-\sqrt{2}B'R_{f_0S'}-\sqrt{2}B''R_{f_0G}|$, 
$|B\cos\theta_a-B'\sin{\theta_a}|$ and 
$|B(R_{f_0NS}+2\sqrt{2}R_{f_0NN})-B'(5R_{f_0N'}+\sqrt{2}R_{f_0S'})-6\sqrt{2}
B''R_{f_0G}|$, 
respectively. 
Adopting the magnitudes for $B$, $B'$ and $B''$ as $2.8{\rm GeV^{-1}}$, 
$12{\rm GeV^{-1}}$ and $7.2{\rm GeV^{-1}}$, respectively, we can get 
the values of the coupling strength for $\phi\to a_0(980)\gamma$, $\phi\to 
f_0(980)\gamma$, $a_0(980)\to\gamma\gamma$ and $f_0(980)\to\gamma\gamma$ decays 
as $2.76{\rm GeV^{-1}}$, $7.15{\rm 
GeV^{-1}}$, $0.84{\rm GeV^{-1}}$ and $6.20{\rm GeV^{-1}}$, respectively. 
The experimental values for these coupling strengths are 
$2.316\pm0.127\ {\rm GeV}^{-1}$, 
$13.8\pm0.54\ {\rm GeV}^{-1}$
$1.137\pm0.214\ {\rm GeV}^{-1}$ and 
$3.920\makebox{$ {+0.591\atop-0.789}$}\ {\rm GeV}^{-1}$, 
then one can say that our model using the VDM and mixing between $qq\bar{q}\bar{q}$ 
and $q\bar{q}$ can explain the radiative decays including $a_0(980)$ 
and $f_0(980)$ consistently. 
%%%%%%%% references %%%%%%%%%%%%%%%%%%%%%%%%%

\end{document}